\documentclass[amsmath,amssymb,aps,10pt,prd,letterpaper,nofootinbib,balancelastpage,notitlepage,superscriptaddress,twocolumn,floatfix]{revtex4-2}
\usepackage{graphicx}	
\usepackage{epstopdf}	
\usepackage{ragged2e}
\usepackage{bm}			
\usepackage[sort&compress]{natbib}	 	
	\setcitestyle{square,numbers,comma}	
\usepackage[colorlinks=true,urlcolor=blue,linkcolor=black,citecolor=red]{hyperref}	
\usepackage{color}
   \definecolor{r}{RGB}{255,0,0}
   \definecolor{p}{RGB}{223,0,225}
   \definecolor{g}{RGB}{0,189,85}
    \definecolor{b}{RGB}{0,171,255}

\newcommand{\figref}[2][]{Fig{#1}.~\ref{fig:#2}}			
\newcommand{\sectref}[2][]{Sec{#1}.~\ref{sect:#2}}		
\newcommand{\appref}[2][x]{Appendi{#1}~\ref{app:#2}}	
\renewcommand{\eqref}[2][]{Eq{#1}.~(\ref{eq:#2})}		
\newcommand{\citeR}[2][]{Ref{#1}.~\cite{#2}}			
\newcommand{\lb}{\ensuremath{\left}}					
\newcommand{\rb}{\ensuremath{\right}}					
\newcommand{\nl}{\nonumber \\ & \quad }					
\newcommand{\bdots}{\ldots\mkern-2mu}
\newcommand{\GeV}{\,\text{GeV}}

\newcommand{\gpgg}{\ensuremath g_{\phi\gamma}}

\begin{document}

\title{Axion Dark Matter Detection with CMB Polarization}
\date{July 30, 2019}
\author{Michael A.~Fedderke}
\email{mfedderke@stanford.edu}
\affiliation{Stanford Institute for Theoretical Physics, Department of Physics, Stanford University, Stanford, CA 94305, USA}
\affiliation{Berkeley Center for Theoretical Physics, Department of Physics, University of California Berkeley, Berkeley, CA 94720, USA}
\affiliation{Theory Group, Physics Division, Lawrence Berkeley National Laboratory, Berkeley, CA 94720, USA}
\author{Peter W.~Graham}
\email{pwgraham@stanford.edu}
\affiliation{Stanford Institute for Theoretical Physics, Department of Physics, Stanford University, Stanford, CA 94305, USA}
\author{Surjeet Rajendran}
\email{surjeet@berkeley.edu}
\affiliation{Department of Physics \& Astronomy, The Johns Hopkins University, Baltimore, MD  21218, USA}
\affiliation{Berkeley Center for Theoretical Physics, Department of Physics, University of California Berkeley, Berkeley, CA 94720, USA}

\begin{abstract}
We point out two ways to search for low-mass axion dark matter using cosmic microwave background (CMB) polarization measurements.
These appear, in particular, to be some of the most promising ways to directly detect fuzzy dark matter.
Axion dark matter causes rotation of the polarization of light passing through it.  
This gives rise to two novel phenomena in the CMB.  
First, the late-time oscillations of the axion field today cause the CMB polarization to oscillate in phase across the entire sky.  
Second, the early-time oscillations of the axion field wash out the polarization produced at last scattering, reducing the polarized fraction (TE and EE power spectra) compared to the standard prediction.  
Since the axion field is oscillating, the common (static) `cosmic birefringence' search is not appropriate for axion dark matter.
These two phenomena can be used to search for axion dark matter at the lighter end of the mass range, with a reach several orders of magnitude beyond current constraints.
We set a limit from the washout effect using existing Planck results, and find significant future discovery potential for CMB detectors searching in particular for the oscillating effect.
\end{abstract}

\maketitle
\tableofcontents

\section{Introduction}
\label{sect:introduction}

Overwhelming gravitational evidence for the existence of dark matter (DM) is to be found over a wide range of astrophysical and cosmological scales (see~\citeR[s]{Bergstrom:2000pn,Bertone:2004pz} for reviews).
Elucidating the non-gravitational properties of the dark matter is one of the most pressing open problems in particle physics. 
A particularly intriguing class of potential DM candidates is supplied by light bosonic degrees of freedom, of which axions and axion-like particles~\cite{Duffy:2009ig,Marsh:2015xka,Graham:2015ouw}, and dark photons~\cite{Holdom:1985ag,Pospelov:2008jk,Chen:2008yi,Redondo:2008ec} are prototypical.

Apart from these being well-motivated degrees of freedom from the standpoint of UV theory (e.g.,~\citeR[s]{Conlon:2006tq,Svrcek:2006yi,Arvanitaki:2009fg,Arvanitaki:2009hb}), the canonical QCD axion is independently extremely well-motivated as a solution to the strong CP problem~\cite{Peccei:1977hh,Peccei:1977ur,Weinberg:1977ma,Wilczek:1977pj}, in certain regions of parameter space.
More generally, much attention in the literature has also recently been directed toward axion-like particles (hereinafter, `axions'),%
\footnote{\label{ftnt:ALPs_vs_axions}%
		Such a particle in general has no fixed relationship between its mass and couplings.
    	We avoid the more elaborate `axion-like particle' nomenclature in favor of `axion' as we refer almost exclusively to such particles in this work; where necessary, we distinguish the `axion' from the `QCD axion', whose mass is fixed once its QCD coupling is specified.
	} %
which interact with the Standard Model via similar couplings to the QCD axion, but which populate a much broader region of mass--coupling parameter space.
In particular, extremely light axions, with astrophysically macroscopic de Broglie wavelengths, can act as `fuzzy dark matter' (FDM) (see~\citeR{Hui:2016ltb} for a review), which may provide a solution to a number of potential small-scale structure anomalies~\cite{Hu:2000ke}.

Light bosonic DM also provides an interesting counterfoil to the long-dominant weakly interacting massive particle (WIMP) paradigm~\cite{Jungman:1995df} for particle dark matter.
Indeed, the absence of definitive evidence for the WIMP in the face of ever-advancing experimental sensitivities~\cite{Angloher:2015ewa,Akerib:2016vxi,Amole:2017dex,TheFermi-LAT:2017vmf,Agnese:2017jvy,Cui:2017nnn,Petricca:2017zdp,Abdallah:2018qtu,Aprile:2018dbl,Abdalla:2018mve,Aprile:2019dbj} is becoming a strong motivation to cast a wider net in the search for the identity of the dark matter.
The very high phase-space occupancy numbers required for light bosons to constitute all of the dark matter~\cite{Hui:2016ltb},%
\footnote{\label{ftnt:production}%
		Over wide regions of parameter space, there are plausible production mechanisms for both axions and dark photons; see, e.g.,~\citeR[s]{Preskill:1982cy,Abbott:1982af,Dine:1982ah,Nelson:2011sf,Graham:2015rva,Agrawal:2018vin,Dror:2018pdh,Co:2018lka,Bastero-Gil:2018uel}.
	} %
resulting in effective classical-field-like behavior of these candidates, gives rise to a broad array of novel effects (see, e.g.,~\citeR{Graham:2013gfa}), requiring experimental approaches quite distinct from the canonical WIMP search techniques.
A plethora of such approaches have recently been proposed; e.g.,~\citeR[s]{Budker:2013hfa,Arvanitaki:2014dfa,Chaudhuri:2014dla,Graham:2015ouw,Kahn:2016aff,TheMADMAXWorkingGroup:2016hpc,Garcon:2017ixh,JacksonKimball:2017elr,Baryakhtar:2018doz,Marsh:2018dlj,Ouellet:2018beu,Arza:2019nta,Bogorad:2019pbu}.

One of the more exotic consequences that arises within the context of light bosonic dark matter is the effect of a pseudoscalar dark-matter axion background field on the propagation of linearly polarized light.
It has long been known that the breaking of parity associated with coupling of any generic non-stationary or non-uniform background pseudoscalar to electromagnetism gives rise to a `birefringence' for the propagation of opposite-helicity photons, manifesting itself as a rotation of the plane of linear polarization of the light by an angle proportional to the difference in the values of the pseudoscalar field at the emission and absorption of the photon~\cite{Carroll:1989vb,Carroll:1991zs,Harari:1992ea,Carroll:1998zi}.%
\footnote{\label{ftnt:no_B_field}%
		This rotation effect is distinct from the axion--photon mixing effects that occur in an external magnetic field background, which also give rise to a rotation effect~\cite{Maiani:1986md,Raffelt:1987im}.
	} %

An extensive literature exists on this subject.
Interferometric searches for the rotation effect have been recently proposed~\cite{DeRocco:2018jwe,Obata:2018vvr,Liu:2018icu,Nagano:2019rbw} for the case of a DM axion in the mass range $m_\phi \sim 10^{-14}$--$10^{-9}$\,eV, which varies on a timescale amenable to a laboratory setting.
Existing searches for the rotation effect when the pseudoscalar varies on temporal or spatial scales inaccessible at Earth- or local space-based facilities exploit the fact that a variety of astrophysical and cosmological sources emit polarized light, which travels over very long baselines to reach Earth, allowing observable net rotations to accumulate over the long travel times.
In the context of polarized emission from astrophysical sources (radio galaxies, pulsars, protoplanetary disks, etc.), searches have encompassed the rotation arising from DM axions~\cite{Alighieri:2010eu,Fujita:2018zaj,Ivanov:2018byi,Liu:2019brz,Caputo:2019tms}, cosmologically slowly varying pseudoscalars~\cite{Carroll:1991zs,Harari:1992ea,Nodland:1997cc,Eisenstein:1997sa,Carroll:1997tc,Leahy:1997wj,Wardle:1997gu,Carroll:1998zi,Alighieri:2010eu,Alighieri:2010pu,Kaufman:2014rpa,Galaverni:2014gca,Whittaker:2017hnz}, and the case where the related Chern--Simons term gives rise to the rotation~\cite{Carroll:1989vb}. 
More germane to the topic of the present work, the rotation of the polarized fraction of the cosmic microwave background (CMB) has been considered in the case where the pseudoscalar field varies slowly on cosmological timescales~\cite{Harari:1992ea,Lue:1998mq,Balaji:2003sw,Feng:2004mq,Feng:2006dp,Liu:2006uh,Ni:2007ar,Pospelov:2008gg,Li:2008tma,Ni:2009fg,Alighieri:2010pu,Caldwell:2011pu,Yadav:2012tn,Li:2013vga,Lee:2013mqa,Galaverni:2014gca,Zhao:2014rya,Zhao:2014yna,Lee:2014rpa,Kaufman:2014rpa,Ni:2014cca,Gubitosi:2014cua,Mei:2014iaa,Gruppuso:2015xza,Gruppuso:2016nhj,Molinari:2016xsy,Gubitosi:2016zdw}; more exotic scenarios have also been considered~\cite{Lepora:1998ix,Choi:1999zy,Feng:2004mq,Li:2008tma,Gubitosi:2009eu,Das:2009ys,Ni:2009fg,Xia:2009ah,Xia:2012ck,Gubitosi:2012rg,Li:2013vga,Zhao:2014rya,Zhao:2014yna,Lee:2014rpa}.%
\footnote{\label{ftnt:grav_probes}%
        Complementarily, the gravitational effects of axions on the CMB power spectra have also been extensively explored (e.g.,~\citeR[s]{Amendola:2005ad,Bozek:2014uqa,Hlozek:2014lca,Schive:2015kza,Urena-Lopez:2015gur,Cedeno:2017sou,Hlozek:2017zzf,Nguyen:2017zqu,Cembranos:2018ulm,Poulin:2018dzj}); these typically probe an axion mass range lighter than we consider in this work, where the axions constitute at most a fraction of the DM (or, for extremely light masses, some fraction of the dark energy).
	} %
Searches by CMB experimental collaborations for either isotropic or anisotropic static `cosmic birefringence' are standard; e.g.,~\citeR[s]{Gluscevic:2009mm,Gluscevic:2012me,Hinshaw:2012aka,Kaufman:2013vbd,Ade:2015cao,Aghanim:2016fhp,Molinari:2016xsy,Array:2017rlf}.

Relatively fewer works have discussed the rotation of the polarization of the CMB due to DM axions~\cite{Finelli:2008jv,Galaverni:2009zz,Liu:2016dcg,Sigl:2018fba,Arias:2012az}, which necessarily vary rapidly on cosmological timescales given that small-scale structure measurements (e.g., Lyman-$\alpha$ measurements~\cite{Viel:2013apy,Irsic:2017yje}, as well as dwarf-galactic structure~\cite{Hu:2000ke}) constrain such axions to have oscillation periods of at most $\mathcal{O}(1\,\text{yr})$.

\section{Executive summary}
\label{sect:ExecutiveSummary}
In this work, we revisit the photon polarization rotation effect arising from a dark-matter axion on the polarized fraction of the CMB.
We point out a number of important phenomena that arise in this context that appear to have either escaped notice or been under-appreciated in previous analyses.

First, we re-emphasize a fundamental point that has long been known in the literature~\cite{Harari:1992ea}: the rotation angle of the plane of polarization that arises from the birefringence induced by the axion--photon coupling is proportional to the difference in the values of the axion field at the emission and the absorption of the photon, and is independent of the details of the behavior of the axion field along the photon trajectory.%
\footnote{\label{ftnt:pseudoscalar}%
	    This statement applies generally to any pseudoscalar field coupled to a photon in a fashion similar to the axion.
	} %

Applying this understanding to the polarized fraction of the CMB, we find two main phenomenological implications:
(1) there will be an AC oscillation, at the axion oscillation period, of the CMB polarization pattern on the sky as measured today, arising as a result of the local DM axion value evolving over the total lifetime of a CMB experiment; and
(2) given that a DM axion must oscillate many times during the CMB decoupling epoch, there is a necessity in the line-of-sight approach to computing CMB anisotropies~\cite{Seljak:1996is} to average the axion-induced linear polarization rotation angle over all possible axion field values explored during the decoupling epoch: this `washes out' the polarization, leading to a reduction of the net polarized fraction of the CMB light as compared to the $\Lambda$CDM expectation. 
Relatedly, the cancellations inherent in having a fast-oscillating axion field at the decoupling epoch result in only a highly suppressed net static (DC) rotation of the CMB linear polarization angle at any point on the sky as compared to a na\"ive estimate of the effect made taking into account only the axion field magnitude at decoupling.

The polarization washout effect is quadratic in the early-time axion field times the axion--photon coupling (i.e., $\gpgg^2 \phi_*^2$, a small parameter), and existing Planck CMB measurements constrain $\gpgg \phi_* \lesssim 1.5\times 10^{-1}$ (future, cosmic-variance-limited reach: $\gpgg \phi_* \lesssim 5.7\times 10^{-2}$). 
Normalizing the axion field $\phi_*$ to be all the dark matter at the decoupling epoch, a $m_\phi \sim 10^{-22}\,$eV dark-matter axion is excluded for $\gpgg \gtrsim 9.6 \times 10^{-14}\GeV^{-1}$ (future: $\gpgg \gtrsim 3.6 \times 10^{-14}\GeV^{-1}$).
The AC effect is linear the late-time axion field times the axion--photon coupling (i.e., $\gpgg \phi_0$, a small parameter); taking an informed estimate for the present detectable amplitude of the AC oscillation to be $0.1^\circ$ (future assumed reach: $0.01^\circ$) allows a reach of $\gpgg \phi_0 \sim 3.5\times 10^{-3}$ (future: $\gpgg \phi_0 \sim 3.5\times 10^{-4}$).
However, the local dark-matter density is smaller than that at decoupling, $\phi_0/\phi_* \sim 10^{-2}$, with the resulting sensitivity to a $m_\phi \sim 10^{-22}\,$eV dark-matter axion from the AC effect being $\gpgg \sim 1.6 \times 10^{-13}\GeV^{-1}$ (future: $\gpgg \sim 1.6 \times 10^{-14}\GeV^{-1}$), comparable to that of the washout effect at current sensitivities.
Both effects have a reach of a few orders of magnitude beyond existing bounds for the lightest possible fuzzy axion dark-matter masses.
The AC effect, being linear in the axion--photon coupling and not cosmic-variance limited, has better potential future reach than the washout effect.

Our work is complementary to, and not in conflict with, the many existing analyses cited in \sectref{introduction} that consider the distinct phenomenology that arises for pseudoscalars that vary slowly on cosmological timescales.
The effects we note, arising from the faster-oscillating DM axions, are largely new (or importantly different from previous discussions of this phenomenology).
We defer a detailed comparison to previous work considering DM axions to the body of the paper.

In the remainder of this paper, we review how the axion-induced modifications to Maxwell's equations give rise to a photon polarization rotation effect (\sectref{ALPElectrodynamics}), which we then employ in a series of increasingly realistic toy models (\sectref{ToyModel}) designed to illustrate the resulting washout and oscillation phenomenology, building up to our analysis of the CMB (\sectref{CMB}) and main results (\figref{result}). 
After discussing past work (\sectref{comparison}), we conclude (\sectref{Conclusion}).
Additional details are given in \appref[ces]{WKB} and \ref{app:stokes}.

\section{Axion electrodynamics}
\label{sect:ALPElectrodynamics}
\label{sect:PolRot}
We consider the action %
\begin{align}
S = \int d^4x \sqrt{-g}\Bigg[ & \frac{1}{2} (\nabla_\mu \phi)(\nabla^\mu \phi) - V(\phi) - \frac{1}{4} F_{\mu\nu} F^{\mu\nu} \nl - J^\mu A_\mu - \frac{1}{4} \gpgg \phi F_{\mu\nu} \widetilde{F}^{\mu\nu} \Bigg],
\label{eq:GRaction}
\end{align}
where $g$ is the metric determinant, $A_\mu$ is the photon, $F_{\mu\nu}$ ($\widetilde{F}_{\mu\nu}$) is the (dual) field-strength tensor, $J^\mu$ is the electromagnetic (EM) current, $\phi$ is the axion, and $\gpgg$ is the axion--photon coupling constant, which has mass-dimension $-1$; throughout this paper, we will assume that $V(\phi) = \frac{1}{2}m_\phi^2 \phi^2$.

It is well known that the axion--photon coupling gives rise to modifications to electrodynamics in an axion field background~\cite{Wilczek:1987mv}. 
As we show in detail in \appref{WKB}, if we specialize to a homogeneous, isotropic Friedmann-Lem\^aitre-Robertson-Walker (FLRW) universe with scale factor $a$, working in the conformal--comoving co-ordinate system $(\eta, \bm{x})$ where $\eta$ is conformal time such that the line element is $ds^2 = [a(\eta)]^2 \lb( d\eta^2 - d\bm{x}^2 \rb)$, and we assume that the axion background field varies along only the $x^3 = z$ spatial direction,%
\footnote{\label{ftnt:notRedshift}%
		In this section, $z = x^3$ is the third spatial coordinate, \emph{not} the redshift.
		} %
$\phi(\eta,\bm{x}) = \phi(\eta,z)$, the photon equations of motion admit the following approximate transverse plane-wave solution (see also, e.g.,~\citeR{Harari:1992ea}, wherein equivalent alternative derivations are presented in terms of the electric and magnetic fields):
\begin{align}
A_0=A_3&=0 \\
A_\sigma(\eta,z) &= A_\sigma(\eta',z')\nl  
    \times \exp\Big[\! -i \omega (\eta-\eta') + ik(z-z') \nl
               \qquad\qquad + i\sigma \frac{\gpgg}{2} \Delta \phi( \eta,z;\eta',z')  \Big] \label{eq:PhaseShift}
\end{align}
with $\omega = k$, where
\begin{align}
\Delta\phi(\eta,z;\eta',z') &\equiv \phi(\eta,z)- \phi(\eta',z') \label{eq:DelPhi}
\end{align}
is the difference in the axion field values between absorption at $(\eta,z)$ and emission at $(\eta',z')$.
Here, we have defined the opposite-helicity transverse photon degrees of freedom 
\begin{align}
A_\sigma \equiv \frac{1}{\sqrt{2}} \lb( A_1 - i \sigma A_2 \rb) \qquad \qquad (\sigma = \pm 1).
\label{eq:AsigmaDefn}
\end{align}
This approximate solution holds in the regime where the axion field varies in space and time much more slowly than the photon field; see \appref{WKB}.

The leading effect of the opposite-sign phase corrections to the two helicity modes $A_\sigma$ shown in \eqref{PhaseShift} is to cause a rotation of the linear polarization of the EM field by an angle $\Delta\theta \propto \gpgg \Delta \phi$, where $\Delta \phi \equiv \phi(\eta_{\text{abs.}},\bm{x}_{\text{abs.}}) - \phi(\eta_{\text{emit}},\bm{x}_{\text{emit}})$ is the difference of the axion field values at photon absorption and photon emission. 

\begin{figure*}[t]
\includegraphics[width=0.75\textwidth]{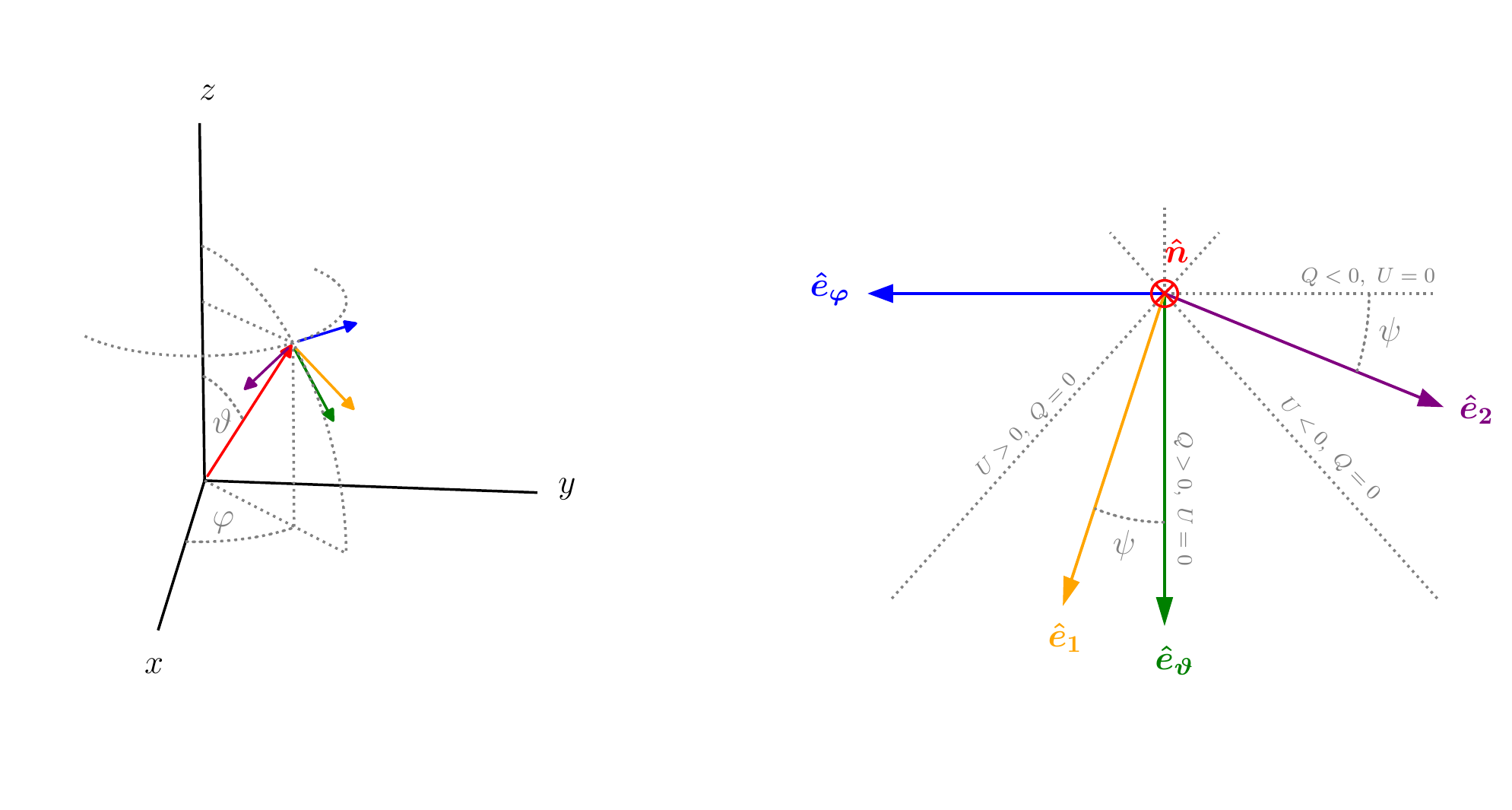}
\caption{ \label{fig:axes} Axis and angle conventions. }
\end{figure*}

Explicitly, the electric field $\bm{E}_\perp \equiv ( E^x , E^y )^{\textsc{t}}$ in the plane transverse to $\bm{\hat{z}}$ undergoes a clockwise rotation as viewed by an observer looking back toward the source of the photon [i.e., an observer directing their view in the $(-\bm{\hat{z}})$-direction]:
\begin{align}
E^i_\perp(\eta,z) &= \lb( \frac{a(\eta')}{a(\eta)} \rb)^2 \exp\lb[ -i \omega (\eta-\eta') + ik(z-z') \rb] \nl \times \bigg( R^{ij}\!\lb[ \frac{\gpgg}{2} \Delta \phi \rb]E^j_\perp(\eta',z') \bigg) \label{eq:Erot0}
\end{align}
with $\omega = k$, where 
\begin{align}
R(\theta) &\equiv \begin{pmatrix} \cos\theta & \sin\theta \\ -\sin\theta & \cos\theta \end{pmatrix},
\label{eq:Erot}
\end{align}
and where the redshift factor correctly accounts for the fact that radiation redshifts in an FLRW universe as $\rho \propto E^2 \propto a^{-4}$.
In \eqref{Erot0}, we have neglected terms $\propto \gpgg \partial_\eta \phi / \omega \ll1$, and we also sum over repeated indices assuming a 3-metric equal to the identity.
The rotation effect is independent of the frequency of the light.

It is worth reiterating that, from \eqref{Erot0}, the $\bm{E}_\perp$-field rotates through an angular excursion
\begin{align}
\Delta \theta &=
\frac{\gpgg}{2} \Delta \phi(\eta_{\text{abs.}},\bm{x}_{\text{abs.}};\eta_{\text{emit}},\bm{x}_{\text{emit}}) \\
&= \frac{\gpgg}{2} \int_{C} ds\, n^\mu\, \partial_\mu \phi\\
&= \frac{\gpgg}{2} \lb[ \phi(\eta_{\text{abs.}},\bm{x}_{\text{abs.}}) - \phi(\eta_{\text{emit}},\bm{x}_{\text{emit}}) \rb] ,
\label{eq:DelPhi2}
\end{align}
where $C$ is the path of the photon in spacetime from the point of emission to the point of absorption, and $n^\mu$ is the null tangent vector to $C$.
This form of the result makes clear that it is indeed a cumulative integrated effect along the whole path of the local derivative axion--photon coupling in \eqref{GRaction}.%
\footnote{\label{ftnt:total_Derivative}%
	        Recall: 
	        \begin{align*}
	            - \frac{1}{4}  \int d^4x \sqrt{-g}\, \gpgg \phi F_{\mu\nu} \widetilde{F}^{\mu\nu} = \frac{1}{2}\int d^4x \sqrt{-g}\, \gpgg (\partial_\mu \phi) A_\nu \widetilde{F}^{\mu\nu}. \end{align*}
		} %

We emphasize strongly that the net rotation effect is \emph{independent of the details of the axion field configuration along the photon path at all points between emission and absorption}: the net rotation depends \emph{only on the initial and final axion field values}~\cite{Harari:1992ea}.
It is immediate from this understanding of the effect that treating the net rotation angle arising from a photon passing through multiple different coherently oscillating axion patches as a stochastic process of random jitters in the polarization angle, as has been done in number of recent works, is incorrect; see further discussion in \sectref{comparison}.
These statements apply whenever the assumptions for using a Wentzel-Kramers-Brillouin (WKB)-like approach to solve the photon equations of motion are satisfied, as detailed in \appref{WKB}; for an axion dark-matter field that is everywhere differentiable and non-relativistic (i.e., with spatial gradients smaller than its temporal gradients), a sufficient condition is that the photon frequency is much larger than the axion mass, $\omega \gg m_\phi$.

\section{Simple toy models}
\label{sect:ToyModel}
In this section we demonstrate the impact of the polarization rotation effect elucidated in \sectref{PolRot} in the context of a series of three related, simplified toy models designed to bear broad similarity to an observer detecting polarized photons emitted from the CMB surface of last scattering.  The third model essentially gives all the physics underlying our results for the effect of a DM axion on the CMB.

\subsection{Minkowski spacetime, source at one instant of time}
\label{sect:ToyModel1}
\newcommand{\nhat}[1][]{\bm{\hat{n}^{#1}}}
\newcommand{\etheta}[1][]{\bm{\hat{e}_{\vartheta}}(\bm{\hat{n}^{#1}})}
\newcommand{\ephi}[1][]{\bm{\hat{e}_{\varphi}}(\bm{\hat{n}^{#1}})}
\newcommand{\ethetaket}[1][]{|\bm{\hat{e}_{\vartheta}}(\bm{\hat{n}^{#1}})\rangle}
\newcommand{\ephiket}[1][]{|\bm{\hat{e}_{\varphi}}(\bm{\hat{n}^{#1}})\rangle}
\newcommand{\eoneket}[1][]{|\bm{\hat{e}_{1}}(\bm{\hat{n}^{#1}})\rangle}
\newcommand{\etwoket}[1][]{|\bm{\hat{e}_{2}}(\bm{\hat{n}^{#1}})\rangle}
To avoid a number of the initially distracting complexities of FLRW spacetime, we first work in Minkowski spacetime with $a\equiv 1$ in this subsection; this implies without loss of generality that we may set $\eta \equiv t$, the cosmic time.
First consider an observer at $x_{\text{obs.}} = (t,\bm{0})$, receiving photons from a distant source localized on the fixed-time surface $x_{\text{source}} = (t',D(t,t')\nhat)$, where $\nhat \equiv \sin\vartheta\cos\varphi \bm{\hat{x}} + \sin\vartheta\sin\varphi \bm{\hat{y}} + \cos\vartheta  \bm{\hat{z}}$ is the direction from observer to source, and the distance $D(t,t') \equiv t - t'$ is such that $x_{\text{obs.}}$ and $x_{\text{source}}$ are light-like separated.
Suppose that in an observation time $\delta t \ll \pi/m_\phi$, the observer receives from the source at random times within that interval a total of $(N+M)$ photons, with $N$ of those photons having been emitted as linearly polarized along the $\bm{\hat{e}_1}$-direction at the source, and $M$ of the photons having been emitted as linearly polarized along the $\bm{\hat{e}_2}$-direction at the source; we denote the quantum-mechanical polarization states of the photons as $\eoneket$ and $\etwoket$, respectively.
We take $\bm{\hat{e}_1} \equiv \cos \psi \,\etheta + \sin\psi \, \ephi$ and  $\bm{\hat{e}_2} \equiv -\cos \psi \, \ephi + \sin\psi \, \etheta$, where $\etheta \equiv \cos\vartheta\cos\varphi \bm{\hat{x}} + \cos\vartheta\sin\varphi  \bm{\hat{y}} - \sin\vartheta  \bm{\hat{z}}$ and $\ephi \equiv -\sin\varphi \bm{\hat{x}} + \cos\varphi  \bm{\hat{y}}$; see \figref{axes}.%
\footnote{\label{ftnt:RHS_system}%
		The sign convention for the $\bm{\hat{e}_{1,2}}$ system is chosen this way such that the triplet $(\bm{\hat{p}} , \bm{\hat{e}_{1}}, \bm{\hat{e}_{2}} )$ forms a right-handed co-ordinate system, where $\bm{\hat{p}}\equiv-\nhat$ is the direction of photon propagation from source to observer.
	} %
We are agnostic here to the mechanism generating the polarization, but assume it to be an immutable, constant characteristic of the source (in particular, we assume the fractions of photons of each polarization in any sized sample of photons do not change).

Assume further that this entire setup occurs in a near-homogeneous background axion field $\phi(t,\bm{x})$ [i.e., $|\partial_t \phi| \gg | \nabla \phi |$]; this implies that $\phi_{\text{obs.}}(t) \equiv \phi(t,\bm{0}) \approx \phi_0 \cos( m_\phi t + \alpha )$ and  $\phi_{\text{source}}(t,t') \equiv \phi(t',D(t,t')\nhat) \approx \phi_0 \cos( m_\phi t' + \beta(t-t') )$, with $\alpha$ and $\beta(t-t')$ being some phases, in general different by $\mathcal{O}(\pi)$ because we will assume that $D$ is much larger than either $\pi/m_\phi$ or the larger axion field coherence length $\sim \pi/(m_\phi v)$.
Note that, as written here, $\beta(t-t') = \beta(D(t-t'))$ is also supposed to capture additional phase variation not made explicit in the $m_\phi t$ term, due to the small spatial gradients of the axion field; as a result of the assumed near-homogeneity, $|\beta(t+ \pi / m_\phi -t') -  \beta(t-t')| \ll \pi$.

Suppose our observer has a polarization-sensitive detector with two sensors, oriented such that one sensor is sensitive to polarization along the $\etheta$-direction, and the other is sensitive along the orthogonal $\ephi$-direction [we denote the photon polarization states along these two axes as $\ethetaket$ and $\ephiket$, respectively].
The observer determines the Stokes parameters (see \appref{stokes}) characterizing the incoming photons by performing an incoherent sum over the photons received from direction $\nhat$ during the observation time $\delta t$:
\begin{align}
I(\nhat) &= \sum_{i=1}^N I_i + \sum_{j=1}^M I_j,\\
V(\nhat) &= 0, \\
Q(\nhat) &=  \sum_{i=1}^N Q_i + \sum_{j=1}^M Q_j,\\
U(\nhat) &= \sum_{i=1}^N U_i + \sum_{j=1}^M U_j,
\label{eq:StokesIncoherentSum}
\end{align}
where $I_{i,j}$, $Q_{i,j}$, and $U_{i,j}$ are the single-photon Stokes parameters defined in \appref{stokesQuantum} with respect to the axes $\etheta$ and $\ephi$; see also~\citeR{Kosowsky:1994cy}.
At the source, the photons are polarized as $| i; (t',D\nhat) \rangle = \eoneket$ for $i=1,\bdots,N$, while $| j; (t',D\nhat) \rangle = \etwoket$ for $j=1,\bdots,M$.
However, we know from \sectref{PolRot} that the linear polarization of each individual photon rotates in a \linebreak (counter-)clockwise direction when $\Delta \phi > 0$ ($<0$), as viewed by an observer looking in the $\nhat$-direction, due to the axion oscillation between photon emission and photon absorption:
\begin{widetext}
\begin{align}
| i; (t,\bm{0})\rangle &= \cos\lb( \frac{\gpgg}{2} \Delta \phi \rb) \eoneket -  \sin\lb( \frac{\gpgg}{2} \Delta \phi \rb) \etwoket \qquad [i=1,\bdots,N]\\[1ex]
			&=  \cos\lb( \psi + \frac{\gpgg}{2} \Delta \phi \rb) \ethetaket  + \sin\lb( \psi + \frac{\gpgg}{2} \Delta \phi \rb) \ephiket \\[1ex]
|j ; (t,\bm{0})\rangle &=  \cos\lb( \frac{\gpgg}{2} \Delta \phi \rb) \etwoket +   \sin\lb( \frac{\gpgg}{2} \Delta \phi\rb) \eoneket \qquad [j=1,\bdots,M] \\[1ex]
	&= - \cos\lb( \psi + \frac{\gpgg}{2} \Delta \phi \rb) \ephiket + \sin\lb( \psi + \frac{\gpgg}{2} \Delta \phi \rb) \ethetaket
\label{eq:ObsETM1}
\end{align}
\end{widetext}
where $\Delta \phi \equiv \phi_0 \lb[ \cos(m_\phi t + \alpha ) - \cos( m_\phi t' + \beta(t-t')) \rb]$.

Therefore, the observer measures the Stokes parameters for the radiation field%
\footnote{\label{ftnt:rot}%
		Note that a right-handed rotation of the axes $(\etheta,\ephi)$ by an angle $\xi$ [i.e., a clockwise rotation of the axes $(\etheta,\ephi)$ as viewed in the right panel of \figref{axes}] would send $\psi \rightarrow \psi - \xi$, with the resulting transformation $Q\pm i U \rightarrow e^{\mp 2i \xi}( Q\pm i U)$.
	} %
\begin{align}
I(\nhat)  &= N+M \\
\qquad V(\nhat) &= 0,\\
(Q\pm i U)(\nhat) &=  \epsilon I \exp\lb[ \pm 2i \lb( \psi + \frac{\gpgg}{2} \Delta\phi \rb) \rb] \\
&= \exp\lb[ \pm 2i \lb( \frac{\gpgg}{2} \Delta\phi \rb) \rb]  (Q\pm i U)_0(\nhat),
\label{eq:StokesTM1}
\end{align}
where we have defined $\epsilon \equiv (N-M)/(N+M)$ as an intrinsic polarization asymmetry of the source, and $(Q\pm i U)_0(\nhat)$ to be the values that would be measured in the axion decoupling limit $\gpgg \rightarrow 0$.

From this setup we can immediately observe one interesting signal: suppose we made two sets of measurements of $I(\nhat),Q(\nhat),U(\nhat)$ at times $t_1$ and $t_2$, separated by something on the order of $|t_1-t_2| \sim \mathcal{O}(\pi /m_\phi)$, using integration durations $\delta t \ll \pi/m_\phi$ for each measurement. 
$I(\nhat)$ is the same at both times and, by virtue of the fact that $\beta$ varies slowly on timescales $\sim \pi/m_\phi$, we can write
\begin{align}
\Delta \phi_2 - \Delta \phi_1 \approx \phi_0 \lb[ \cos(m_\phi t_2 + \alpha ) - \cos(m_\phi t_1 + \alpha ) \rb].
\label{eq:DelPhi1}
\end{align}
Therefore, $Q(\nhat)$ and $U(\nhat)$ are observed to rotate into each other as a function of $t_2$ (for fixed $t_1$) as though the detector angle $\psi$ were being oscillated with a period $\sim 2\pi/m_\phi$ and excursion amplitude $\gpgg\phi_0/2$.  
In other words, we see an axion-induced `AC oscillation' of the $Q(\nhat)$ and $U(\nhat)$ Stokes parameters measured to be emitted from a light-like separated source located on a fixed-time surface in the past, assuming that the source properties are otherwise fixed over the relevant timescales, and the source has some intrinsic net polarization asymmetry.

\subsection{Minkowski spacetime, source smeared in time}
\label{sect:ToyModel2}
The toy model of \sectref{ToyModel1} fails to account for a situation in which the source is not localized at one point in time on the past lightcone of the observer; we extend the model to that case in this subsection.

We continue in Minkowski spacetime with $a=1$, taking $\eta \equiv t$.
Suppose now that our source is otherwise unchanged from the toy model of \sectref{ToyModel1}, but that there is instead a smeared region along the past lightcone of the observer which contributes all the photons arriving at the observer within a time interval $\delta t$ around the time $t$. 
That is, instead of being localized at $\bm{x} = r\nhat = D\nhat$ at one fixed instant of time $t'$, it is rather the case that our source has some unit-normalized probability density function (PDF) $g(t')$ to be spatially located at $\bm{x}_k = r_k\nhat = D_k \nhat= (t-t_k')\nhat$ at the time $t'=t_k'$ when it emits each of the $k=1,\bdots,N+M$ photons that the observer receives in the $\delta t$-interval around time $t$.%
\footnote{\label{ftnt:clarify_setup_2}%
		Note that $( x_k - x_{\text{observer}} )^2 = ( t - t_k)^2 - D_k^2 = 0$; the photons are emitted at slightly different times \emph{and distances}, such that they all arrive at the same instant at the observer.
	} %
We assume $g(\tilde{t}')$ has non-negligible support only in the region $\tilde{t}' \in [ t' - T/2 , t' + T/2 ]$ where $t-t' \gg T \gg \pi/m_\phi$: i.e., the duration $T$ of the photon emission / source smearing is small compared to the distance to the source, but large compared to the axion period (had we instead assumed the hierarchy $t-t' \gg \pi/m_\phi \gg T$, such that the axion field oscillates slowly during the period of photon emission, we would recover the result of \sectref{ToyModel1}).
We additionally assume that $2\pi |\partial_t g(t)| \ll m_\phi$, so that $g(t)$ is fairly constant on one axion period.

In this second case, the expressions for the observer-measured Stokes parameters are modified to account for the fact that each individual photon has its own value of $\Delta \phi$:
\begin{align}
I &= N+M,\\[1ex]
V &= 0, \\
(Q \pm i U)(\nhat) &=  \sum_{i=1}^N \exp\lb[ \pm 2 i \lb( \psi + \frac{\gpgg}{2} \Delta\phi_i \rb) \rb] \nl- \sum_{j=1}^M \exp\lb[ \pm2 i \lb( \psi + \frac{\gpgg}{2} \Delta\phi_j \rb) \rb],
\label{eq:StokesTM2}
\end{align}
where 
\begin{align}
\Delta \phi_{k}  \equiv \Delta \phi(t,t_{k}') 
&=\phi_0 \Big[ \cos(m_\phi t + \alpha ) \nl\qquad\! - \cos( m_\phi t_k' + \beta(t-t_k')) \Big],
\end{align}
with $t_k'$ the emission time for photon $k$.
For fixed $t$, and assuming that we are in the limit where $N$ and $M$ are large enough that we may pass to the continuum limit (i.e., we have large photon statistics), we have
\begin{align}
&(Q \pm i U)(\nhat) \nl
= \epsilon I \int d\tilde{t}'\, g(\tilde{t}') \exp\lb[ \pm 2i \lb(\psi +  \frac{\gpgg}{2} \Delta \phi(t,\tilde{t}')  \rb) \rb].
\label{eq:StokesTM2b}
\end{align}
Now, 
\begin{align}
2 \psi + \gpgg \Delta \phi(t,\tilde{t}') &= 2\psi +  \gpgg \phi_0 \cos(m_\phi t + \alpha ) \nl - \gpgg \phi_0 \cos( m_\phi \tilde{t}' + \beta(t-\tilde{t}'));
\label{eq:phase}
\end{align}
although $\beta(t-\tilde{t}')$ can vary by $\mathcal{O}(\pi)$ on timescales $\sim T$, we still have the scale separation $|m_\phi T | \gg |\beta(t-t' - T/2)- \beta(t-t' + T/2)|$.
We can thus separate the integration domain into a large number of subdomains $\tilde{t}' \in [ \tilde{t}'_n - \delta \tilde{t}'/2 , \tilde{t}'_n + \delta \tilde{t}' / 2 ]$, where $\delta \tilde{t}' = 2\pi/m_\phi$, such that $|\beta(t-\tilde{t}'_n - \delta t'/2)- \beta(t-\tilde{t}'_n +\delta t'/2)| \ll \pi$, so that we may take $ \beta(t-\tilde{t}') \approx  \beta(t-\tilde{t}'_n) \equiv \beta_n$ to be independent of $\tilde{t}'$ in each subdomain. 
Moreover, we have assumed that $g(\tilde{t}')$ varies slowly on the timescale $\delta \tilde{t}'$, and so may approximate $g(\tilde{t}') \approx g(\tilde{t}_n') \equiv g_n$ in each subdomain:
\begin{widetext}
\begin{align}
(Q\pm i U)(\nhat) &\approx \epsilon I  \sum_n g_n \int_{\tilde{t}'_n-\delta t'/2}^{\tilde{t}'_n + \delta t'/2} d\tilde{t}' \exp\lb[ \pm 2i \lb(\psi +  \frac{\gpgg}{2} \phi_0 \cos(m_\phi t + \alpha ) - \frac{\gpgg}{2} \phi_0\cos( m_\phi \tilde{t}' + \beta_n ) \rb) \rb] \label{eq:Qexpr} \\[1ex]
& = \epsilon I  \lb( \sum_n g_n \delta \tilde{t}' \rb) \exp\lb[ \pm 2 i \lb(\psi +  \frac{\gpgg}{2} \phi_0 \cos(m_\phi t + \alpha ) \rb) \rb] J_0 \lb( \gpgg\phi_0 \rb) \\[1ex]
&= \epsilon I \lb( \int d\tilde{t}' g(\tilde{t}') \rb) \exp\lb[ \pm 2 i \lb(\psi +  \frac{\gpgg}{2} \phi_0 \cos(m_\phi t + \alpha ) \rb) \rb] J_0 \lb( \gpgg\phi_0 \rb)  \\[1ex]
&= J_0 \lb( \gpgg \phi_0 \rb) \exp\lb[ \pm 2 i \lb( \frac{\gpgg}{2} \phi_0 \cos(m_\phi t + \alpha ) \rb) \rb] (Q\pm i U)_0(\nhat),
\label{eq:StokesTM2c}
\end{align}
\end{widetext}
where $(Q\pm i U)_0$ are as before the values that would be measured in the axion decoupling limit $\gpgg\rightarrow 0$.
We also used that $g(\tilde{t})$ is a unit-normalized PDF, and the integral representation of the Bessel function of the first kind, $J_0(x)$: $\int_{-\pi}^{\pi}dx \exp\lb[ i A + i B \cos( x + \delta ) \rb] = 2\pi e^{iA} J_0(B)$.%
\footnote{\label{ftnt:J0expansion}%
		$J_0(x) \approx 1-x^2/4 + \cdots$ for $x \ll 1$.
	} %

We thus see again that there is an `AC effect' ascribable to the axion field oscillation at the observer: a time-varying rotation of $Q(\nhat)$ and $U(\nhat)$ into each other as though the detector were being rotated about the $\nhat$-axis through an angular excursion of amplitude $\gpgg \phi_0/2$ at period $2\pi/m_\phi$. 

We also now find a new effect, `polarization washout', ascribable to the axion field oscillating rapidly at the source: the `smearing' of the source in time leads to a smearing of polarizations measured in any arbitrary fixed observer frame (provided the source is not varying), leading to an effective reduction in the polarized fraction interpreted to be produced at the source: $\epsilon \rightarrow \epsilon  J_0(\gpgg \phi_0)$. 
We note that this second effect is only valid in the regime in which the emission does not occur during a period rapid compared to the axion oscillation frequency (i.e., that we have $T \gg 2\pi/m_\phi$); in the opposite limit ($T \ll 2\pi/m_\phi$), the results of \sectref{ToyModel1} would apply instead.

Finally, we emphasize that, in this regime, there is no unsuppressed leading-order static (DC) rotation of the polarization axis that arises from the axion field behavior at the source~\cite{Arvanitaki:2009fg,Marsh:2015xka}, in sharp contrast to the usual treatment of static cosmic birefringence which assumes a setup closer to our initial case from \sectref{ToyModel1}, with $T \ll  2\pi/m_\phi$.
We do however note that we have in our above derivation neglected the `last partially uncanceled oscillation' which would lead to a residual overall rotation from source to observer that is power-law suppressed by $\sim (m_\phi T)^{-1} $ compared to the na\"ive estimate of an $\mathcal{O}( \gpgg \phi_0 )$ rotation (note that we disagree on the size of the suppression compared to~\citeR{Arvanitaki:2009fg}, which claims an exponential $\sim e^{-m_\phi T}$ suppression).

If we imagined the extension of the preceding analysis to a collection of sources at different angular positions $\nhat$ with respect to the observer, and all beaming photons toward the observer, and analyzed how the polarization from each location would be impacted, our conclusions would hold point-by-point: there is a $Q(\nhat)\leftrightarrow U(\nhat)$ oscillation driven by the axion field at the location of the observer, and a second-order reduction in the $Q(\nhat)$ and $U(\nhat)$ values arising from the axion field oscillating many times during the emission from the smeared source.

\subsection{FLRW spacetime, source smeared in time}
\label{sect:ToyModel3}
Consider the third and final toy model, which is taken to be similar to the model in \sectref{ToyModel2}, with two exceptions: (a) the analysis is conducted in an FLRW spacetime instead of in Minkowski spacetime, and (b) instead of assuming near-homogeneity of the axion field across our entire setup, the axion field at the source is taken to be approximately the axion field as it would be at the epoch of CMB last scattering, while the axion field at the observer is taken to be the axion field as it would be locally in the Milky Way (MW).  

Since the analysis of \sectref{PolRot} (or, more specifically, the detailed derivation in \appref{WKB}) indicates that the polarization rotation effect operates in FLRW the same as it would in Minkowski, (a) is an almost trivial change: we simply locate the observer at $x_{\text{obs.}} = ( \eta , \bm{0} )$ and have the source for each of the $(N+M)$ photons located at $x_{\text{source},\, n}=(\eta'_n,D_n\nhat)$, where $D_n \equiv D(\eta,\eta'_n) \equiv \eta - \eta'_n$ is the co-moving distance between the observer and source, and $\eta$ is conformal time. 
We assume that $\eta'_n$ is drawn from the distribution $g(z(\eta'))$ with $g(z)$ the visibility function, which we take to be localized around $z = z_*$, where $z_*$ is the redshift of last scattering.%
\footnote{\label{ftnt:z_is_redshift}%
		In contrast to \sectref{PolRot}, here $z$ is the redshift, \emph{not} the third spatial co-ordinate, $x^3$.
	} %

To evaluate the axion field at the source, we consider the axion field equation, \eqref{ALPEOM}, ignoring the source term on the RHS (see discussion in \appref{WKB}), and assume that spatial gradients are irrelevant: $|\nabla \phi|/a \ll | \partial_t \phi| $, which is justified by our assumption that the axion field serves as the cold, non-relativistic dark matter.
Moreover, we evaluate the axion equation of motion in the FLRW spacetime describing our Universe:
\begin{align}
\hat{\phi}(t) &\equiv \lb[ a(t) \rb]^{3/2} \phi(t) \\
\Rightarrow \ddot{\hat{\phi}}(t) &+ \lb[ m_\phi^2 - \frac{3}{2} \lb\{ \frac{\ddot a}{a} + \frac{1}{2} \lb( \frac{\dot a }{a} \rb)^2 \rb\} \rb] \hat{\phi}(t) =0 \\
\ddot{\hat{\phi}}(t) &+ m_\phi^2 \lb[ 1  + \frac{6\pi}{m_\phi^2 M_{\text{Pl.}}^2} p_{\text{total}} \rb] \hat{\phi}(t) = 0,
\label{eq:phiEoM2}
\end{align}
where $\dot{} \equiv \partial_t$, we defined $\hat{\phi}(t)$ to factor off the dominant scale-factor dependence of $\phi(t)$, and in the last step we used the Friedmann equations to trade out derivatives of $a$ for the total pressure from all sources: 
\begin{align}
p_{\text{total}} &= p_{\text{matter}} +  p_{\Lambda} + p_{R} \\
                &= p_{\Lambda} + p_{R} \\
				&= - \rho_\Lambda + \frac{1}{3} \rho_R \\
				&=  \rho_c \lb[ - \Omega_\Lambda^0 + \frac{\Omega_R^0}{3} a^{-4} \rb],
\label{eq:ptotal}
\end{align}
where $\Omega_i^0 \equiv \rho_i(t_0)/\rho_c$ are the present-day fractional energy densities, and we used that the matter fields (including the cold DM axion field at this epoch) do not contribute to the pressure, as their equation of state is $w\approx 0$.
But then using $\Omega_R^0 = \Omega_M^0 / ( 1+z_{\text{eq.}} )$, we have
\begin{align}
\frac{6\pi}{m_\phi^2M_{\text{Pl.}}^2} p_{\text{total}} = \lb( \frac{3 H_0}{2m_\phi} \rb)^2 \lb[ - \Omega_\Lambda^0 + \frac{\Omega_M^0}{3} \frac{(1+z)^{4}}{1+z_{\text{eq.}}} \rb].
\label{eq:ptotal2}
\end{align}
Using the latest Planck (TT\,+\,TE\,+\,EE\,+\,lowE\,+\,lens-\linebreak ing\,+\,BAO) results~\cite{Aghanim:2018eyx}, $\Omega_\Lambda^0 = 0.6889(56) $, $\Omega_M^0 = 0.3111(56)$, $z_{\text{eq.}} = 3387(21)$, $z_{*} = 1089.89(21)$, and $H_0 = 67.77(42) \,\text{km}\,\text{s}^{-1}\,\text{Mpc}^{-1}$, we find that 
\begin{align}
&\frac{6\pi}{m_\phi^2M_{\text{Pl.}}^2} p_{\text{total}} \nl
\approx  \Bigg[ - 3.3\times 10^{-24} + 2.1\times 10^{-16} \lb( \frac{1+z}{1+z_{*}}\rb)^4 \Bigg]\nl\quad \times \lb( \frac{m_\phi}{10^{-21}\,\text{eV}} \rb)^{-2}.
\label{eq:ptotal3}
\end{align}
As we are assuming in this section that $z\sim z_*$ within $\sim 20\%$, the $[ \, \cdots ]$-bracket in \eqref{phiEoM2} is negligibly different from 1, and we can write
\begin{align}
\phi(t',\nhat) &\approx \phi_*(\nhat)  \lb[ \frac{1+z'}{1+z_*} \rb]^{3/2} \cos\lb( m_\phi t' + \beta \rb),
\label{eq:phiFLRW}
\end{align}
where $\phi_*(\nhat)$ is a normalization that in principle we allow to vary across the sky, $\beta$ is a phase, and $z' = z(t')$ is the redshift.
If the axion constitutes a fraction $\kappa$ of all the cold dark matter at the average redshift of decoupling, and we ignore the small matter perturbations at decoupling which would give rise to variations of $\phi_*(\nhat)$ across the sky, we have 
\begin{align}
\frac{1}{2} m_\phi^2 \phi_*^2 &= \kappa \rho_c \Omega_c^0 (1+z_*)^3 \nonumber \\
                &= (3 \kappa /8\pi) M_{\text{Pl.}}^2 \Omega^0_ch^2 (H_0/h)^2 (1+z_*)^3 \label{eq:phistar0} \\[1ex] 
\Rightarrow \phi_*(\nhat) &= 1.6 \times10^{11}\,\text{GeV} \times \kappa^{1/2} \nl\times  \lb(\frac{m_\phi}{10^{-21}\,\text{eV}}\rb)^{-1} \times \lb(\frac{ \Omega^0_c h^2}{ 0.11933 } \rb)^{1/2},
\label{eq:phistar}
\end{align}
where we used the Planck result $\Omega_c^0 h^2 = 0.11933(91)$~\cite{Aghanim:2018eyx}.

In evaluating the axion field at the observer, we use the axion field local to us in the MW: 
\begin{align}
\phi(t) \approx \phi_0 \cos( m_\phi t + \alpha(t) ),
\label{eq:localALP}
\end{align}
where $\alpha(t)$ is a phase approximately constant on timescales shorter than the axion coherence time $t_{\text{coh.}}\sim2\pi/(m_\phi v_0^2)$ with $v_0$ the MW virial velocity, and we take $\phi_0$ to be normalized such that the axion field constitutes a fraction $\kappa$ of the local DM density, $\rho_0 \approx 0.3 \, \text{GeV}/\text{cm}^3$:
\begin{align}
\frac{1}{2} m_\phi^2 \phi_0^2 &= \kappa \rho_0 \label{eq:phi00} \\
\Rightarrow \phi_0 &= 2.1 \times10^{9}\,\text{GeV} \times \kappa^{1/2} \nl \times \lb( \frac{m_\phi}{10^{-21}\,\text{eV}}\rb)^{-1} \times \lb( \frac{\rho_0}{ 0.3\, \text{GeV}/\text{cm}^3} \rb)^{1/2}.
\label{eq:phi0}
\end{align}
The reader will note that the axion field normalizations given at Eqs.~(\ref{eq:phistar}) and (\ref{eq:phi0}) differ by only two orders of magnitude, which is understood by virtue of the fact that $\rho \propto \phi^2$, our Galaxy is locally overdense compared to the present-day critical density ($\rho_c \sim 4.8\times10^{-6}\, \text{GeV}/\text{cm}^3$) by roughly 5 orders of magnitude, and the dark matter in the Universe around last scattering was $\sim (1+z_*)^3\sim10^9$ times denser that the present-day critical density; it is important that these two normalizations, $\phi_*$ and $\phi_0$, are not wildly dissimilar.

In our Universe, the photon visibility function $g(z)$ has a FWHM spanning $z \sim 1000$--$1200$~\cite{Dodelson2003,Weinberg2008}, corresponding to a cosmic time interval $T= \Delta t \sim 10^5$\,yrs, while an axion with mass $m_\phi$ oscillates with a period $T_{\phi} \sim 0.1 \,\text{yrs} \times( 10^{-21}\,\text{eV} / m_\phi )$; $g(z)$ thus varies slowly on the timescale of the axion oscillation in the parameter region of interest.
As a result, most of the analysis of \sectref{ToyModel2} now goes through as before, with the exception that we need to additionally note that in a time period $\delta\tilde{t} = 2\pi/m_\phi$ around $z=z_*$, $\tilde{z} \equiv z(\tilde{t}')$ is such that we have $\delta \tilde{z} \equiv z(\tilde{t}'+\delta\tilde{t}') - z(\tilde{t}') \ll \tilde{z}$, so that at the step analogous to the subdivision of the integration domain just above \eqref{Qexpr}, we may write $1+\tilde{z} \approx 1+\tilde{z}_n$ in each subdomain:
\begin{widetext}
\begin{align}
&(Q \pm i U)(\nhat)\nonumber \\
&\approx \epsilon I \int d\tilde{t}' g(\tilde{t}') \exp\lb[ \pm 2i \lb(\psi +  \frac{\gpgg}{2} \phi_0 \cos(m_\phi t + \alpha ) - \frac{\gpgg}{2} \phi_*(\nhat)  \lb[ \frac{1+\tilde{z}}{1+z_*} \rb]^{3/2} \cos( m_\phi \tilde{t}' + \beta(t,\tilde{t}') ) \rb) \rb] \\
&\approx \epsilon I  \sum_n g_n \int_{\tilde{t}'_n-\delta t'/2}^{\tilde{t}'_n + \delta t'/2} d\tilde{t}' \exp\lb[ \pm 2i \lb(\psi +  \frac{\gpgg}{2} \phi_0  \cos(m_\phi t + \alpha ) - \frac{\gpgg}{2} \phi_*(\nhat) \lb[ \frac{1+\tilde{z}_n}{1+z_*} \rb]^{3/2}  \cos( m_\phi \tilde{t}' + \beta_n ) \rb) \rb]\\
& = \epsilon I \exp\lb[ \pm 2i \lb(\psi +  \frac{\gpgg}{2} \phi_0 \cos(m_\phi t + \alpha ) \rb) \rb] \lb[ \sum_n g_n \delta \tilde{t}'\, J_0 \lb( \gpgg \phi_*(\nhat)  \lb[ \frac{1+\tilde{z}_n}{1+z_*} \rb]^{3/2} \rb) \rb] \\
&=  \epsilon I \exp\lb[ \pm 2i \lb(\psi +  \frac{\gpgg}{2} \phi_0 \cos(m_\phi t + \alpha ) \rb) \rb] \int d\tilde{t}'\, g(\tilde{t}')\, J_0 \lb( \gpgg \phi_*(\nhat)  \lb[ \frac{1+\tilde{z}}{1+z_*} \rb]^{3/2} \rb)  \\
&=  \epsilon I \exp\lb[ \pm 2i \lb(\psi +  \frac{\gpgg}{2} \phi_0 \cos(m_\phi t + \alpha ) \rb) \rb] \int d\tilde{z}\, g(\tilde{z})\, J_0 \lb( \gpgg \phi_*(\nhat)  \lb[ \frac{1+\tilde{z}}{1+z_*} \rb]^{3/2} \rb)  \\
&= J_0 \lb[ \gpgg \langle\phi_*\rangle(\nhat) \rb] \exp\lb[ \pm 2i \lb( \frac{\gpgg}{2} \phi_0 \cos(m_\phi t + \alpha ) \rb) \rb] (Q\pm i U)_0(\nhat), 
\label{eq:ToyModel3Result}
\end{align}
\end{widetext}
where $(Q\pm i U)_0$ are as before the values that would be measured in the axion decoupling limit $\gpgg \rightarrow 0$, and where for convenience we defined $g(\tilde{t}')\equiv g(\tilde{z}) ( d\tilde{z} / d\tilde{t}' )$, as well as a visibility-function-weighted average axion field amplitude at decoupling:
\begin{align}
&J_0 \lb[ \gpgg\langle \phi_* \rangle(\nhat)  \rb] \nonumber \\
&\equiv \int d\tilde{z}\, g(\tilde{z})\, J_0 \lb( \gpgg \phi_*(\nhat) \lb[ \frac{1+\tilde{z}}{1+z_*} \rb]^{3/2} \rb)\label{eq:AverageDefn0} \\
&\approx J_0 \lb[ \gpgg \phi_*(\nhat)  \rb],
\label{eq:AverageDefn}
\end{align}
where the latter approximate equality holds since $g(\tilde{z})$ is fairly strongly peaked around $\tilde{z}=z_*$.
Within the context of this most-developed toy model, we see the same overall effects as in \sectref{ToyModel2}, with the magnitude of the AC oscillation effect now explicitly seen to be governed by the amplitude of the local axion field value, and the magnitude of the washout effect controlled by a visibility-function-weighted average of the axion field amplitude at last scattering.
 
Finally, in the limit where $\gpgg \phi_{0,*} \ll 1$, so that all the axion effects are small, \eqref{ToyModel3Result} immediately tells us that the effect of $\phi_*$ manifests itself only quadratically in the reduction of the Stokes $Q(\nhat)$ and $U(\nhat)$ parameter amplitudes ($J_0(x) \sim 1 - x^2/4$ for small $x$), while the effect of $\phi_0$ appears linearly in the amplitude of the oscillating $Q(\nhat)\leftrightarrow U(\nhat)$ rotation at period $2\pi/m_\phi$.
Therefore, despite the hierarchy $\gpgg \phi_0\sim 10^{-2} \gpgg \phi_*\ll 1$, it is incumbent upon us in what follows to keep track of both the washout and the AC effect, as it is not \emph{a priori} clear how the magnitudes of $(\gpgg\langle\phi_*\rangle)^2$ and $\gpgg\phi_0$ compare in the relevant part of parameter space.
In other words, even though the axion field is larger at last scattering than today, the washout effect is quadratic while the AC effect is linear in the axion field, so both effects could be important.

\begin{figure*}[t]
\includegraphics[width=0.85\textwidth]{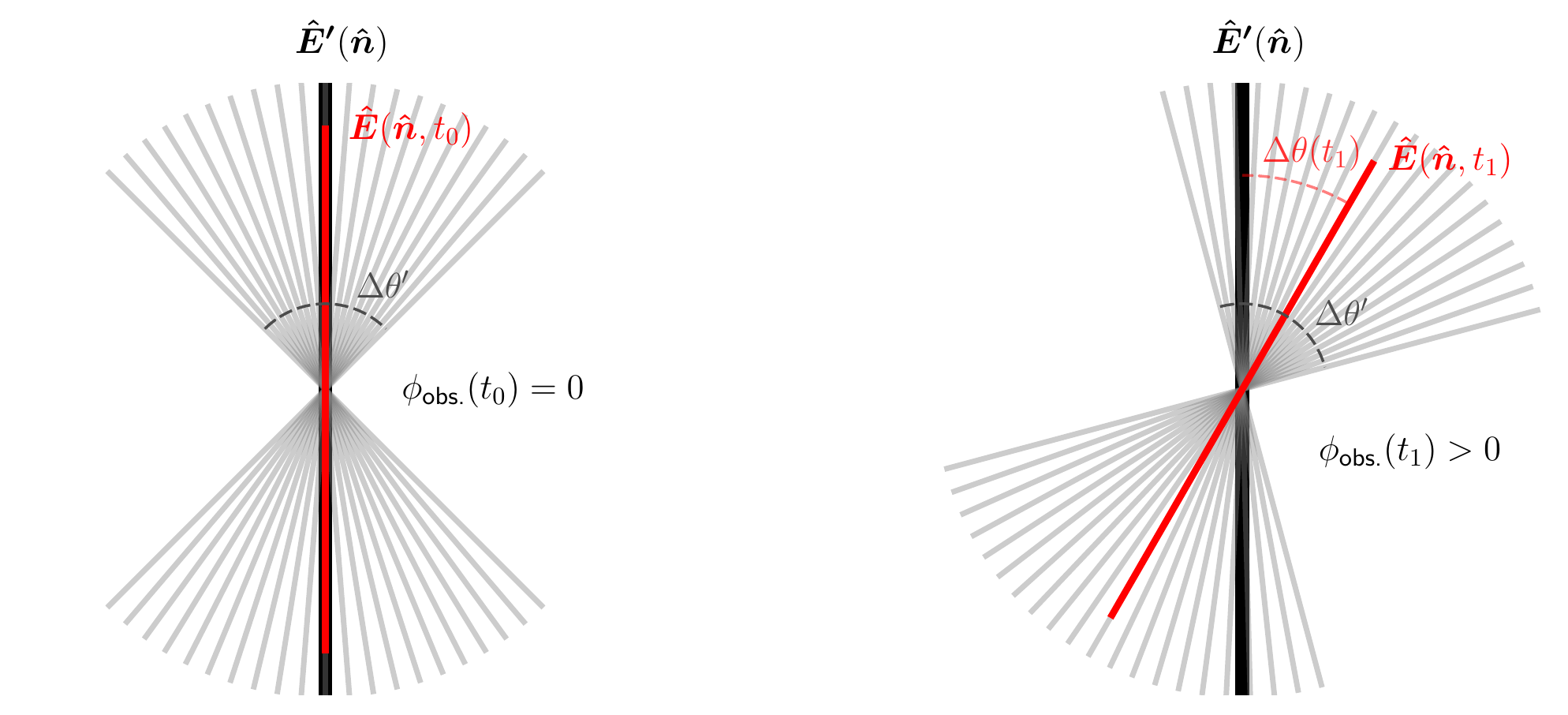}
\caption{ \label{fig:fan} Heuristic explanation of the washout and AC effects. Effects exaggerated for clarity; see text for further discussion. 
		In both panels, the thick black lines marked $\bm{\hat{E}'}(\nhat)$ indicate the axis of polarization favored by the local CMB temperature anisotropies, at some point on the sky, $\nhat$. 
		The thin grey lines are a representative sample of the `fanned-out' (by an angle $\Delta \theta' \sim \gpgg \phi_*$) set of incoming photon polarization directions seen by an observer at Earth, arising as a result of the individual photons being emitted at random points during the axion field evolution around decoupling.
		The thick red lines marked $\bm{\hat{E}}(\nhat,t_i)$ indicate the average of these fanned-out polarizations detected at Earth; the reduced length of those lines is supposed to indicate the reduction in observed polarization amplitude as compared to that at the source.
		In the left panel, we consider an observer making the measurement at a time $t_0$ such that the axion field at the observer is instantaneously zero [$\phi_{\text{obs.}}(t_0)=0$], so there is instantaneously no offset between the direction of the polarization intended to be imprinted by CMB source conditions, and the direction of the polarization observed at Earth: $\bm{\hat{E}'}(\nhat) \parallel \bm{\hat{E}}(\nhat,t_0)$. 
		That is, we instantaneously see only the washout.
		In the right panel, we consider an observer making its measurement at time $t_1$ some fraction of an axion period later [$\phi_{\text{obs.}}(t_1)>0$], such that the entire polarization pattern at the detector also undergoes an overall rotation $\Delta \theta(t_1)$ compared to the direction intended to be imprinted by CMB source conditions: $\bm{\hat{E}'}(\nhat) \nparallel \bm{\hat{E}}(\nhat,t_1)$.
		As the axion field at the observer continues to evolve in time, the angular offset $\Delta \theta(t)$ of the pattern will oscillate back and forth.
		Note that nothing in this figure is intended to imply a net unsuppressed static DC rotation offset of the received polarization pattern; these are simply snapshots in time of the AC oscillation $\Delta \theta(t)$ of the polarization angle offset.
		}
		\end{figure*}

To conclude this section, we give a simple heuristic, pictorial understanding for the origins of the washout and AC effects; see \figref{fan}.
For any fixed direction $\nhat$ on the sky, the local temperature quadrupole conditions prevailing at decoupling set up some preferred polarization direction, $\bm{\hat{E}}'(\nhat)$, at that point. 
In the absence of an axion-induced rotation effect, this same polarization direction would be seen by an observer at or near Earth.
However, suppose we turn on the axion field, and focus initially on a fixed observation time at which the local axion field value at the detector is momentarily zero (\figref{fan}, left panel).
An observer at Earth receives photons which were emitted at random times during any one of the $\gtrsim 10^5$ axion oscillation periods during the decoupling epoch; these photons thus fully and randomly sample all possible axion field values explored around the decoupling epoch.
Because the polarization of each such photon therefore ends up rotating between emission and absorption by a different amount (set by the axion field value at the emission time of each photon), the observer sees instead a series of incoming photons whose polarizations are `fanned out' over an angular range $\Delta \theta' \sim \gpgg \phi_*$ around the direction $\bm{\hat{E}}'(\nhat)$. 
But the observer (i.e., a bolometric CMB detector) naturally performs an incoherent sum over the fanned-out collection of incoming photons, and as a result detects only their average, $\bm{\hat{E}}(\nhat,t_0)$, as the incoming polarization from direction $\nhat$.
As can be seen from the left panel in \figref{fan}, in the limit of large photon statistics, this averaging merely results in a reduction in the amplitude of the polarization, but no net static rotation.
Finally, as we allow time to evolve at the observer and the local axion field evolves away from zero (\figref{fan}, right panel), the entire `fan pattern' further executes an angular oscillation of amplitude $\Delta \theta \sim \gpgg \phi_0$, which naturally also oscillates the average of the fanned-out polarizations, $\bm{\hat{E}}(\nhat,t)$; this is the AC effect.

\section{CMB observables and limits}
\label{sect:CMB}
By construction, the toy model results of \sectref{ToyModel3} give a good approximation to the effects to be expected in the realistic CMB case:%
\footnote{\label{ftnt:noVpol}%
		It is necessary to remark that, in our toy models, we dictated that only linear polarization was present, so the elliptical polarization Stokes parameter $V$ was necessarily zero.
		For the CMB observations we consider in this work, $V=0$ is a good assumption.
		Only linear polarization is generated primordially in the standard picture of Thompson scattering of photons off unpolarized electrons in the plasma around the decoupling epoch~\cite{Dodelson2003,Weinberg2008}.
		Elliptical polarization can be generated either in non-standard scenarios (e.g.,~\citeR[s]{Alexander:2008fp,Giovannini:2010ar}) or by foreground effects (e.g.,~\citeR[s]{De:2014qza,Ejlli:2018ucq,Ejlli_2018}).
		The CMB measurements we consider explicitly in this work \cite{Aghanim:2018eyx} were taken in the $\mathcal{O}(100\,\text{GHz})$ frequency range, where strong upper limits on the CMB elliptical polarization component exist (e.g.,~\citeR{Nagy:2017csq}). 
		The effects of foregrounds or non-standard physics in generating elliptical polarization \emph{at these frequencies} are thus necessarily small, and it is justified to approximate $V=0$ throughout this paper when discussing the effects of the axion--photon interaction on the primordial linear polarization measured at these frequencies.
		Note however that effects generating $V\neq 0$ are in general frequency dependent (e.g., \citeR[s]{Ejlli:2018ucq,Ejlli_2018}); consideration of the impact of the axion--photon interaction on measurements of the CMB polarization at frequencies outside the $\mathcal{O}(100\,\text{GHz})$ region may thus need to consider $V\neq0$ for full generality.
	} %
a polarization power washout on all (pristine; see comments below) scales, and an AC oscillation effect at period $2\pi/m_\phi$, with the size of the effects given by \eqref{ToyModel3Result}.
As the polarization rotation effect we are considering  impacts only the propagation, and not (at least at leading order) the production, of polarized photons, it is not necessary at the level of accuracy required for the present work to modify any publicly available CMB codes (e.g., \textsc{cmbf}ast~\cite{Seljak:1996is} \textsc{camb}~\cite{Lewis:1999bs}) to explicitly track the washout effect through the decoupling epoch.
At high $l$ (i.e., $l\gtrsim 20$), we judge that simply taking the usual output of unmodified CMB codes, and transforming the output polarizations in the manner indicated by \eqref{ToyModel3Result} is sufficient.
At low $l$, where the primary CMB anisotropies have been significantly reprocessed by the reionized universe, this approach would not be correct; we discuss this case in more detail below.

Furthermore, note that the visibility function $g(z)$ appears only in the average $\langle \phi_* \rangle$, and our results are thus insensitive to its exact shape, so long as it is strongly peaked near $z\sim z_*$, but slowly varying on the timescale of one axion oscillation; in the relevant region of axion mass parameter space, this is always the case with any physically reasonable $g(z)$.

While the washout effect can be constrained using current publicly available CMB power spectrum data (from, e.g., Planck~\cite{Aghanim:2018eyx}), a correct treatment of the AC oscillation effect necessarily requires access to non-public time-series CMB polarization data, and demands a more rigorous treatment of experimental systematics than is appropriate for the scope of the present work.
Therefore, while a combined approach considering both effects simultaneously would be preferable, we instead present two independent estimates for the reach of each effect: (a) the current reach for a washout-only search using the sensitivity of current Planck~\cite{Aghanim:2018eyx} CMB power spectra, as well as a projection for future reach; and (b) the approximate reach attainable by a detailed time-series analysis as informed by the statistical power of current DC searches for cosmic birefringence, as well as a minimal projection for possible future reach.

\begin{figure*}[!p]
\includegraphics[width=0.9\textwidth]{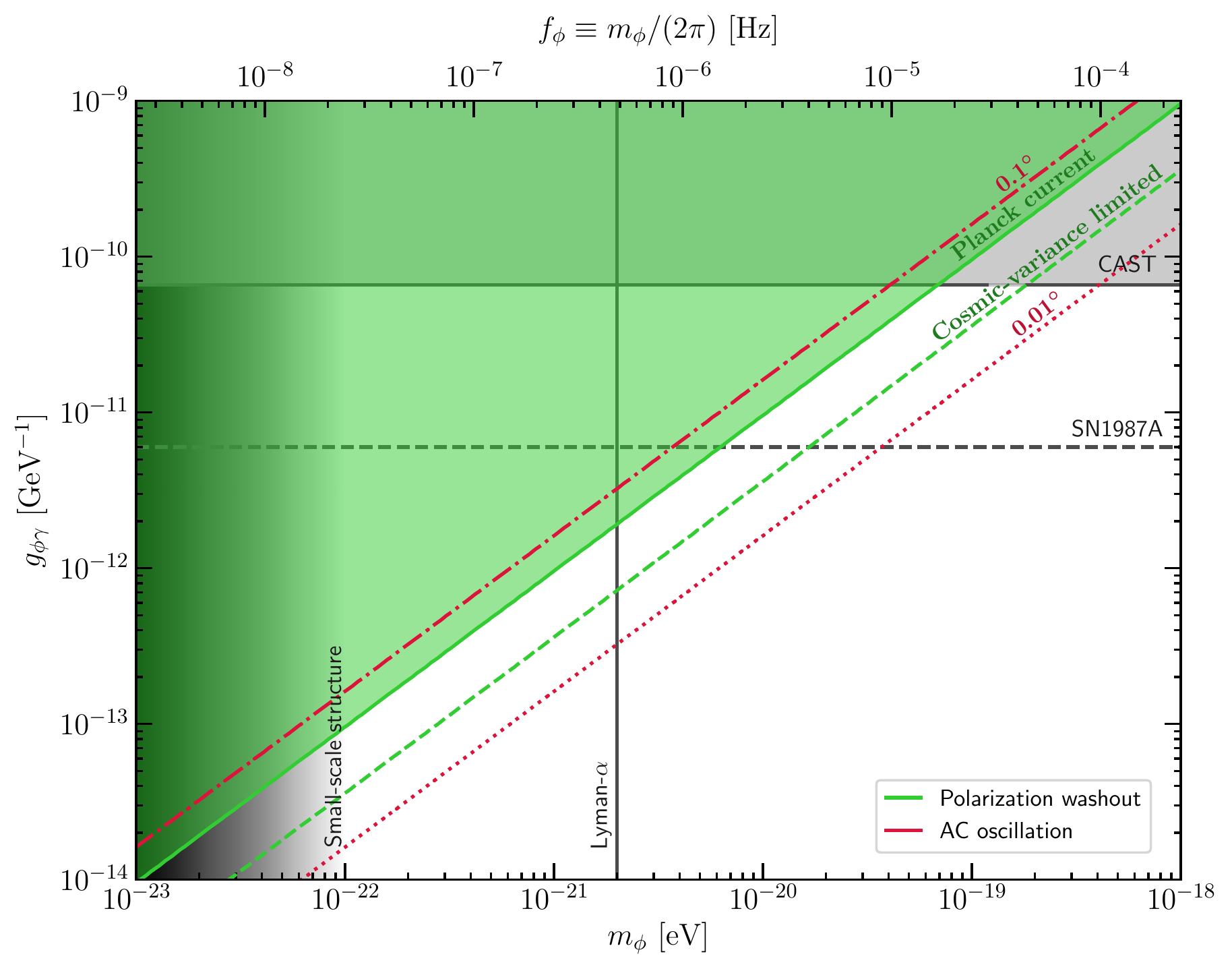}
\caption{ \label{fig:result}
	        Reach for current and future searches for the washout and AC oscillation effects induced by an axion field of mass $m_\phi$ constituting all of the dark matter ($\kappa = 1$, in the notation of \sectref{ToyModel3}). 
		We find that the diagonal (green) shaded region is excluded on the grounds that a reduction in the predicted polarization amplitude of at least $0.58\%$ is in conflict at $95\%$ confidence with Planck measurements~\cite{Aghanim:2018eyx}; see \sectref{currentFull}.
		The diagonal dashed (green) line indicates how the bound would improve if the 95\%-confidence upper limit on the fractional reduction were improved to 0.082\%, our projection for the sensitivity if all three of the power spectra $C^{TT,TE,EE}_l$ were measured to the cosmic-variance limit out to $l_{\text{max}}=3000$ (assuming the Planck sky coverage, $f_{\text{sky}} = 0.577$); see \sectref{future}.
		The diagonal dash-dotted (red) line indicates where the AC oscillation effect has an amplitude of rotation of at least $0.1^\circ$; based on the statistical power of the DC search conducted in~\citeR{Aghanim:2016fhp}, we estimate that, at a minimum, this region may be excludable at 95\% confidence given current sensitivity of the Planck measurements, although this would require a dedicated time-series analysis to confirm.
		The diagonal dotted (red) line indicates the reach of the AC oscillation search if a $0.01^\circ$ rotation amplitude were detectable, as an example of future potential.
		For comparison, the reach for the AC search exceeds the current (respectively, future) washout reach for a rotation amplitude of $0.059^\circ$ ($0.022^\circ$).
		Note that all of our bounds and projections are upward-sloping diagonal lines $\gpgg \propto m_\phi$ because all the rotation effects are $\propto \gpgg \phi_{0,*}$, while $m_\phi^2 \phi_{0,*}^2 \propto \text{const.}$; see \eqref[s]{phistar0} and (\ref{eq:phi00}).
		The vertical (grey) line marked `Lyman-$\alpha$' denotes the smallest axion mass claimed to be consistent  at 95\% confidence with observed small-scale structure in the Lyman-$\alpha$ forest~\cite{Irsic:2017yje}.
		The graduated shaded (grey) region labeled `small-scale structure' is where the axion de Broglie wavelength is large enough to cause tension with observed galactic small-scale structure~\cite{Hui:2016ltb}.
		The horizontal (grey) shaded region is the current 95\%-confidence exclusion limit from CAST~\cite{Anastassopoulos:2017ftl}.
		The horizontal dashed (grey) line is a limit arising from conversion to gamma-rays in galactic magnetic fields of axions emitted by SN1987A~\cite{Payez:2014xsa} (we take their `$10.8M_{\odot}$, Jansson and Farrar' limit).
		}
\end{figure*}

\subsection{Washout of polarization}
\label{sect:washout}
For the purposes of the washout analysis, we neglect the local oscillating axion field, and consider the washout effects via their impact on the CMB $TT$, $TE$, and $EE$ auto- and cross-correlation power spectra.

In considering full-experiment combined maps produced by, e.g., Planck, we are implicitly averaging over the \emph{local} axion phase evolution too, as each point on the sky was revisited multiple times over the multi-year duration of the Planck experiment (see, e.g., footnote 34 of~\citeR{Akrami:2018vks}) and such repeat observations were obviously combined without making any corrections for an actual on-the-sky variation of the CMB polarization arising in the manner we consider in this work.
However, since the local axion field amplitude is noticeably smaller than the field amplitude at decoupling, $\phi_0 \sim 10^{-2} \phi_*$, this second averaging / washout effect should not dramatically alter the prediction for the polarization reduction which we obtain considering only the axion field behavior at the decoupling epoch.

In harmonic-space all-sky CMB analysis (see, e.g.,~\citeR[s]{Zaldarriaga:1996xe,Kamionkowski:1996ks}), the observed temperature maps $T(\nhat)$ are decomposed over a basis of scalar spherical harmonics, and the observed $Q(\nhat)$ and $U(\nhat)$ maps are decomposed over a basis of spin-weight-2 spherical harmonics:
\begin{align}
T(\nhat) &\equiv \sum_{l,m} a_{T,lm} Y_{lm}(\nhat)\\
(Q\pm i U)(\nhat) &\equiv \sum_{l,m} a_{\pm2,lm}\times {}_{\pm2}Y_{lm}(\nhat),
\label{eq:AllSkyMaps}
\end{align}
from which the $E$ and $B$ modes are constructed as~\cite{Zaldarriaga:1996xe}
\begin{align}
a_{E,lm} &\equiv - \frac{1}{2} \lb( a_{+2,lm} + a_{-2,lm} \rb) \\
a_{B,lm} &\equiv \frac{i}{2} \lb( a_{+2,lm} - a_{-2,lm} \rb),
\label{eq:AllSkyMultipoles}
\end{align}
with the observed $C_{XY,l}$ auto- and cross-correlation power spectra being defined as 
\begin{align}
C^{\text{obs.}}_{XY,l} &\equiv \frac{1}{2l+1} \sum_{m}  a_{X,lm}^* a_{Y,lm}  ,
\label{eq:PowerSpectra}
\end{align}
for $X,Y \in \{ T,E,B \}$.
Up to the limitations imposed by cosmic variance, the observed $C^{\text{obs.}}_{XY,l}$ are compared to the true $C_{XY,l} \equiv \langle C^{\text{obs.}}_{XY,l}\rangle$ (with $\langle\, \cdots \rangle$ denoting the average over the ensemble of possible statistical realizations of the sky), which are a prediction of the cosmology of interest.

From this, it is clear that an axion-induced reduction in the amplitude of $Q(\nhat)$ and $U(\nhat)$ point-by-point on the sky results in a reduction in the amplitudes of the $C_{TE,EE,BB}$ power spectra compared to the values expected for these quantities in a base $\Lambda$CDM cosmology with the amplitude of $C_{TT}$ held fixed.
Quantitatively, the relevant effects are [assuming that $\langle\phi_*\rangle(\nhat) \approx \langle\phi_*\rangle$ is independent of $\nhat$, except for small fluctuations]:
\begin{align}
(Q\pm i U)(\nhat) &\rightarrow J_0 \lb( \gpgg \langle\phi_*\rangle  \rb) (Q\pm i U)(\nhat) \\
\Rightarrow  a_{\{E,B\},lm} &\rightarrow J_0 \lb( \gpgg \langle\phi_*\rangle \rb) a_{\{E,B\},lm},
 \end{align}
implying that%
\footnote{\label{ftnt:noCVC}%
		Note that since the temperature and polarization maps are imperfectly correlated~\cite{Coulson:1994qw,Seljak:1996ti}, it is appropriate to treat the effects at the lower of the power spectra, and in particular to incorporate cosmic-variance uncertainties on all the power spectra.
	} %
 \begin{align}
C_{TE,l} &\rightarrow J_0 \lb( \gpgg \langle\phi_*\rangle \rb) C_{TE,l}, \\
C_{EE,l}  &\rightarrow \lb[ J_0 \lb( \gpgg \langle\phi_*\rangle  \rb) \rb]^2 C_{EE,l},
\label{eq:TEEEReduction}
\end{align}
where $\langle \phi_* \rangle$ is as defined at \eqref{AverageDefn0}.
For values of $\gpgg \langle\phi_*\rangle$ such that $J_0\lb( \gpgg \langle\phi_*\rangle  \rb)=1-\delta$ for some $\delta\ll 1$, it follows that the fractional reduction in the $EE$ power spectrum, $|(\Delta C_{EE,l})/C_{EE,l}| \sim 2\delta$, is approximately twice the fractional reduction in the $TE$ power spectrum, $|(\Delta C_{TE,l})/C_{TE,l}| \sim \delta$.

Note, however, that the approach outlined above is only correct for those `pristine' CMB multipoles that have not been significantly re-processed since the decoupling epoch; this prescription fails at low $l$, since the polarization power for $l \lesssim 20$ is significantly contaminated by polarization generated by rescattering after reionization~\cite{Hu:1997hv}.
We therefore mostly confine our attention to the high-$l$ multipoles of the CMB power spectra, which reflect the polarization imprinted at the last scattering surface; where necessary, we include low-$l$ data, but make an approximation as to how the effect operates there (see \sectref{currentFull}).

A fully correct treatment of our effect at low-$l$ would require running a full set of Boltzmann evolution equations to incorporate also the axion field oscillation during the reionization era.
This is however computationally challenging, owing to the extreme ratio of scales in our problem: the timescale of variation of the axion field, $T_\phi \lesssim 1$\,year in our parameter range, is orders of magnitude less than other relevant timescales (the time since reionization began, length of the decoupling epoch, the age of the Universe, etc.), which makes the resulting system of equations stiff.
While usual treatments of faster-oscillating axion fields (see, e.g., \citeR{Hlozek:2014lca}) have utilized an effective fluid description for the axions that essentially averages over the axion oscillations \cite{Turner:1983he}, all our polarization effects arise explicitly from the axion oscillations; we would not be able to follow such an approach to alleviate the computational burden.
As the uncertainties of measured power spectra are dominated by (large) cosmic variance at low $l$, most of the statistical power of our analyses does not arise from the lowest few multipole bins.
Therefore, the extra computational burden associated with a fully correct treatment at low $l$ would not be justified within the scope of the present work, and we discuss in more detail below how we treat the low-$l$ data approximately, when necessary.

Relatedly, although we do not take this approach, we caution the reader that the numerics of computing the CMB polarizations with modified versions of standard Boltzmann codes requires some care, even at high $l$: when the axion field oscillates rapidly on cosmological timescales, an accurate numerical evaluation of the line-of-sight time integrals over the polarization source functions (see, e.g.,~\citeR{Seljak:1996is}, and modifications in~\citeR[s]{Finelli:2008jv,Galaverni:2009zz,Liu:2006uh,Gubitosi:2014cua}) requires a much finer time sampling than would ordinarily be required in the case of standard $\Lambda$CDM, in order not to alias over the rapid axion oscillations.

In the remainder of this section, we detail our analysis of the current and future constraining power of CMB data on the washout effect, making use of the full-mission Planck~\cite{Aghanim:2018eyx} results for the measured power spectra, along with the current best-fit $\Lambda$CDM model (specifically, the Planck TT+TE+EE+lowE+lensing fit~\cite{Aghanim:2018eyx}).\\

\subsubsection{Current bound: approximate analysis}
\label{sect:currentApprox}
We first perform an approximate analysis, considering only $50 \leq l \leq l_{\text{max}}$ to avoid the low-$l$ reionization bump ($l_{\text{max}}= 1996$ is the highest available Planck polarization multipole~\cite{Aghanim:2018eyx}).
In this multipole range, there is within $\Lambda$CDM models a single undetermined overall normalization, $A \propto A_se^{-2\tau}$, which governs the $C_{\{TT,TE,EE\},l}$ spectra~\cite{Dodelson2003,Weinberg2008}; this is the most obvious $\Lambda$CDM parameter combination whose variation could \emph{partially} compensate the polarization washout effect.%
\footnote{\label{ftnt:partialCompensation}%
        To develop an intuitive understanding for why this compensation can occur, take the \emph{crude} approximation $u(C_l)\propto C_l$; each term in the sum over $l$ in \eqref{like1} would then take the approximate form $\sim \lb(1-A\rb)^2 + \lb(1-(1-\delta)A\rb)^2+\lb(1-(1-\delta)^2A\rb)^2$ where $\delta \approx r^2/4$ for small $r$, if exact agreement between the observed and theory $C_l$ is additionally crudely assumed. 
        If $A=1$ is fixed, this term contributes an amount $\sim 5\delta^2$ to the sum; on the other hand, if $A$ is profiled over and allowed to float to $A=1+\delta$, then the contribution to the sum is only $\sim 2\delta^2$.
        Failure to profile over $A$ would result in setting a bound on $r$ a factor of $\sim (5/2)^{1/4} \approx 1.25$ too strong (although note that this heuristic argument should not be taken seriously in any quantitative sense).
        } %
Therefore, we profile over the joint overall normalization $A$ of the three power spectra in performing a one-parameter profile likelihood ratio test for the alternative hypothesis that assumes the existence of a non-zero polarization reduction parameter $r \equiv \gpgg \langle \phi_*(m_\phi) \rangle$, with $\langle \phi_*(m_\phi) \rangle \approx \phi_*(m_\phi) $ as in \eqref{phistar}, against the null hypothesis of no polarization reduction, $r=0$.
Quantitatively, we define%
\footnote{\label{ftnt:OffDiagonalSmall}%
		We make the approximation that we may ignore any off-diagonal elements in the covariance matrix in constructing this likelihood.
	} %
\begin{widetext}
\begin{align}
- 2 \ln \mathcal{L}( r \equiv \gpgg \langle \phi_*(m_\phi) \rangle ; A) &\equiv  \sum_{XY \in \{TT,TE,EE\} }\sum_{l=50}^{l=l_{\text{max}}} \lb( \frac{ C_{XY,l}^{\text{observed}} - A\times f_{XY}(r)\times C_{XY,l}^{\text{theory}} }{ u( C_{XY,l}^{\text{observed}} ) } \rb)^2 \label{eq:like1}
\end{align}
\end{widetext}
where
\begin{align}
f_{XY}(r) &\equiv \begin{cases} 1 & XY = TT \\ J_0 \lb( r\rb) & XY = TE \\  \lb[ J_0 \lb( r \rb) \rb]^2 & XY = EE \end{cases}, \label{eq:like2}
\end{align}
and where $l_{\text{max}}=1996$, $ C_{XY,l}^{\text{observed}} $ is the Planck observed power, $C_{XY,l}^{\text{theory}}$ is the best-fit (TT+TE+EE+lowE+lensing) prediction of the base $\Lambda$CDM model, and $u( C_{XY,l}^{\text{observed}} )$ is the uncertainty in the $C_{XY,l}^{\text{observed}}$ (all as quoted by~\citeR{Aghanim:2018eyx}).
We form the test statistic 
\begin{align}
\Delta \chi^2( r ) \equiv - 2 \ln \lb( \frac{ \mathcal{L}( r ; \hat{\hat{A}}( r ) ) }{ \mathcal{L}( r=0 ; \hat{A}(r=0) ) } \rb),
\label{eq:test-stat}
\end{align}
where hatted quantities are the best-fit values of parameters given a fixed value of $r$.%
\footnote{\label{ftnt:Ar0varies}%
		By not fixing $\hat{A}(r=0)\equiv1$, we allow for the possibility in our procedure that the differences between our simplified version of the Planck likelihood in this approximate analysis and the full Planck analysis pipeline give rise to a small offset between the parameter point that our procedure deems to be the best-fit cosmology and the parameter point that the full Planck analysis pipeline selects as the best-fit cosmology.
		This offset is at the per mille level; i.e., $\hat{A}(r=0)-1\sim\mathcal{O}(10^{-3}$). 
	} %
The one-parameter 95\%-confidence limit is then obtained as $\Delta \chi^2(r_{95} ) = 3.84$; or, expressed in terms of a bound on $\gpgg \phi_*(m_\phi)$:
\begin{align}
1 - J_0 \lb( \gpgg \phi_*(m_\phi) \rb) &\leq 4.9 \times 10^{-3} \quad [\text{approximate}] \label{eq:CMBLimitWashout0}
\end{align}
[i.e., a 0.49\% reduction in $(Q\pm i U)$], which implies
\begin{align}
\gpgg &\lesssim 8.8 \times  10^{-13}\,\text{GeV}^{-1} \qquad\quad [\text{approximate}] \nl \times \lb(\frac{m_\phi}{10^{-21}\,\text{eV}}\rb)\times \lb(\kappa \times \frac{ \Omega^0_c h^2}{ 0.11933 } \rb)^{-1/2},\label{eq:CMBLimitWashout}
\end{align}
where in the second line we used the small-argument expansion of the Bessel function.

\subsubsection{Current bound: full analysis}
\label{sect:currentFull}
The exclusion limit at \eqref{CMBLimitWashout} from the analysis procedure of \sectref{currentApprox} is approximate.
In order to give a rigorous exclusion limit from the current Planck data, we undertook a more complete analysis in which we profiled over the full six-dimensional $\Lambda$CDM base parameter set $\bm{\Theta} \equiv (\Omega_bh^2,\,\Omega_ch^2,\,100\,\theta_{\textsc{mc}},\,\tau,\,\ln(10^{10}A_s),\,n_s)$.
We form the test statistic for the one-parameter profile likelihood ratio test on $r$ as%
\footnote{\label{ftnt:Theta0varies}%
		We again permit $\hat{\bm{\Theta}}(r=0)$ to float to account for differences in our analysis procedure and the full Planck analysis pipeline causing a small shift in what we deem the best-fit cosmology as compared to the Planck results.
	} %
\begin{align}
\Delta \chi^2( r ) \equiv - 2 \ln \lb( \frac{ \mathcal{L}( r ; \hat{\hat{\bm{\Theta}}}( r ) ) }{ \mathcal{L}( r=0 ; \hat{\bm{\Theta}}(r=0) ) } \rb),
\label{eq:test-stat2}
\end{align}
utilizing \textsc{camb}~\cite{Lewis:1999bs} to compute the necessary power spectra, and sampling the $\Lambda$CDM parameters to find the best-fit parameters using Monte Carlo (MC) techniques. 
As this is computationally intensive, we were informed in our choices of the values of $r$ to investigate by the results of our approximate analysis of \sectref{currentApprox}.

Moreover, excluding the low-$l$ multipoles completely in this full analysis would lead to an almost unconstrained parameter degeneracy between $A_s$ and $\tau$.
In this part of our analysis we therefore retain all multipoles down to $l=2$, but in order to not incorrectly wash out the polarization of the non-pristine multipoles at $l\lesssim 20$, we replace the function $f_{XY}(r)$ as defined at \eqref{like2} by $f_{XY}(r) \rightarrow (1-h(l)) + f_{XY}(r) h(l)$, where $h(l) \equiv \frac{1}{2}\lb[ 1 + \tanh((l-l_0)/\delta l) \rb]$ with $l_0 = 20$ and $\delta l = 5$; this approximate choice only turns the polarization washout effect on for $l\gtrsim 20$, and it breaks the parameter degeneracy in the same sense that the low-$l$ reionization bump usually breaks this parameter degeneracy in base $\Lambda$CDM models.
While this is an approximation, most of the statistical power in the limits does not arise from this range of multipoles (where cosmic variance is large), so the impact of this approximation on our results is not significant.

This more rigorous analysis procedure yields a one-parameter 95\%-confidence exclusion bound given the current Planck data of 
\begin{align}
1 - J_0 \lb( \gpgg \phi_*(m_\phi) \rb) &\leq 5.8 \times 10^{-3} \label{eq:CMBLimitWashout0Full}
\end{align}
[i.e., a 0.58\% reduction in $(Q\pm i U)$], or
\begin{align}
 \gpgg &\lesssim 9.6 \times  10^{-13}\,\text{GeV}^{-1}  \nl \times \lb(\frac{m_\phi}{10^{-21}\,\text{eV}}\rb) \times \lb(\kappa \times \frac{ \Omega^0_c h^2}{ 0.11933 } \rb)^{-1/2}.\label{eq:CMBLimitWashoutFull} 
\end{align}
The bound shown by the diagonal (green) shaded region in \figref{result} reflects the result at \eqref{CMBLimitWashoutFull}.
This full result reflects weakening of the limit on $\gpgg$ at fixed $m_\phi$ by only $9\%$ as compared to the approximate bound at \eqref{CMBLimitWashout}.

Our full analysis thus validates the approximate approach of \sectref{currentApprox} within 10\%, and also definitively settles the question as to whether any multi-parameter combinations of base $\Lambda$CDM parameters can completely mimic or compensate our effect: no such combinations exist.

\subsubsection{Future reach}
\label{sect:future}
In order to give a perhaps optimistic estimate where this search could reach in the future, we assume that the $TT$, $TE$, and $EE$ power spectra can all be measured at the cosmic-variance limit from $l = 50$ to $l_{\text{max}} = 3000$ with the same sky coverage as Planck, $f_{\text{sky}} = 0.577$~\cite{Aghanim:2018eyx}.
At large $l$, the results of~\citeR{Percival:2006ss} indicate that the likelihood for observing the data $\bm{\hat{C}}_l \equiv ( \hat{C}_l^{TT} , \hat{C}_l^{TE} , \hat{C}_l^{EE} )$ given a model predicting $\bm{C}_l \equiv ( C_l^{TT} , C_l^{TE} , C_l^{EE} )$ is approximately a multivariate Gaussian with the covariance matrix%
\footnote{\label{ftnt:diag_cov_mat}%
		We have also checked that the covariance matrix may be approximated by its diagonal entries with essentially negligible impact on the estimated future bound.
		Note however that, since the $TE$--$TE$ diagonal entry of $\Sigma_l$ is not just proportional to $(C_{l}^{TE})^2$, one must still always still run, e.g., \textsc{camb}~\cite{Lewis:1999bs} to obtain an accurate value for the ratio $C_{l}^{TT}C_{l}^{EE} / (C_l^{TE})^2$. 	
	} %
\begin{align}
\Sigma_l &\approx \frac{1}{(2l+1)f_{\text{sky}}} \nl \times \begin{pmatrix} 2 (C_l^{TT})^2 & 2 C_{l}^{TT} C_{l}^{TE} & 2 (C_l^{TE})^2 \\[1ex]
											2 C_{l}^{TT} C_{l}^{TE} & C_l^{TT} C_l^{EE} + ( C_l^{TE} )^2 & 2 C_l^{TE} C_l^{EE} \\[1ex]
											2 (C_l^{TE})^2	& 2 C_l^{TE} C_l^{EE} & 2 ( C_l^{EE} ) ^2 \end{pmatrix},
\label{eq:CovMat}
\end{align}
in the limit where cosmic variance dominates noise.
For our estimate for future sensitivity, we revert to the approximate analysis procedure of \sectref{currentApprox}, having verified in \sectref{currentFull} that it is accurate within 10\% for setting exclusion bounds.
As we are setting a future projected exclusion bound on $r \equiv \gpgg \langle \phi_*(m_\phi) \rangle$, we take as our `data' the predictions of the Planck TT+TE+EE+lowE+lensing best fit~\cite{Aghanim:2018eyx} base $\Lambda$CDM model \emph{assuming no polarization washout}: $\bm{\hat{C}}_l \equiv (\bm{C}_l)_{\text{Planck }\Lambda\text{CDM best fit}}$ (which we re-compute using \textsc{camb}~\cite{Lewis:1999bs}).%
\footnote{\label{ftnt:Asimov}%
	    This is the `Asimov data set'~\cite{Cowan:2010js} for this exclusion bound.
	} %
The `theory' we fit to these data, $\bm{C}_l$, is taken to be the modified $\Lambda$CDM prediction \emph{with washout} that was used at \eqref[s]{like1} and (\ref{eq:like2}): $C^{XY}_l \equiv A \times f_{XY}(r)\times (C^{XY}_l)_{\text{Planck }\Lambda\text{CDM best fit}}$.
Our likelihood function for the one-parameter profile likelihood ratio test on $r$ is thus
\begin{align}
-2 \ln \mathcal{L}( r \equiv \gpgg \langle \phi_*(m_\phi) \rangle ; A ) &\equiv \sum_{l=50}^{l_{\text{max}}} \bm{\Delta}_l^T \Sigma_l^{-1} \bm{\Delta}_l,
\end{align}
where
\begin{align}
\bm{\Delta}_l &\equiv \bm{C}_l - \bm{\hat{C}}_l\\
\Rightarrow \Delta_l^{XY} &\equiv (C^{XY}_l)_{\text{Planck }\Lambda\text{CDM best fit}} \nl \times \lb[ A \times f_{XY}(r) - 1 \rb]. 
\label{eq:Like3}
\end{align}
Performing the same profile likelihood ratio test as contemplated at \eqref{test-stat}, we find that our estimate for a future 95\%-confidence exclusion bound under these assumptions is given by
\begin{align}
1 - J_0 \lb( \gpgg \phi_*(m_\phi) \rb) &\leq 8.2 \times 10^{-4} \label{eq:WashoutEstFut0} 
\end{align}
[i.e., a 0.082\% reduction in $(Q\pm i U)$], or
\begin{align}
\gpgg &\lesssim 3.6 \times  10^{-13}\,\text{GeV}^{-1}  \nl \times \lb(\frac{m_\phi}{10^{-21}\,\text{eV}}\rb) \times \lb(\kappa \times \frac{ \Omega^0_c h^2}{ 0.11933 } \rb)^{-1/2},\label{eq:WashoutEstFut1} 
\end{align}
which is a factor of $\sim\sqrt{7}$ better than the current limit given at \eqref{CMBLimitWashoutFull}.
The diagonal dashed (green) line in \figref{result} shows the result at \eqref{WashoutEstFut1}.
Of course, if higher multipoles can be probed at the cosmic-variance limit, this reach could potentially be improved further; however, significant foregrounds exist for the temperature maps above $l \sim 3000$ (see, e.g.,~\citeR{Abazajian:2016yjj}), so sensitivity estimates are more complicated to make.

\subsubsection{Further discussion}
\label{sect:furtherDiscussion}
In all of the above, we have neglected the small fluctuations in the value of $\langle \phi_* (\nhat) \rangle$ across different locations in the sky owing to fluctuations in the DM density; these would induce only a very small (fractionally $\sim 10^{-5}$) modulation on top of the uniform amount by which the amplitudes of $C_{TE,EE,BB}$ are reduced from their base $\Lambda$CDM predicted values.
Additionally, the approximation $\langle \phi_* \rangle \approx \phi_*$ is valid to within (at worst) 10\%, as the visibility function is sufficiently sharply peaked around $z\sim z_*$.

Finally, as we noted in \sectref{ToyModel3}, there is no unsuppressed net (static DC) rotation of the polarization at any point on the sky induced by the axion field at last scattering; there will only be a residual `last uncanceled period' effect (i.e., an incomplete cancellation of the positive and negative polarization rotation angle deviations averaged over the CMB decoupling epoch), and this will be suppressed by a factor of $\sim \Delta t_{\text{osc.}} / \Delta t_{\text{decoupling}} \lesssim 10^{-5}$.
Thus, the usual birefringence searches for a conversion of CMB polarization $E$ modes into $B$ modes are not very useful to search for a dark-matter axion since the field oscillates rapidly at the decoupling epoch; see also comments in this regard in~\citeR[s]{Finelli:2008jv,Galaverni:2009zz}.
Of course, there is still an overall AC polarization rotation due to the late-time effect of the axion field today (at the position of the detector).  We analyze this in the next subsection.

\subsection{AC oscillation of polarization}
\label{sect:AC}
In considering the AC oscillation effect, we ignore the power reduction from the previous subsection.

Searches for the $Q(\nhat) \leftrightarrow U(\nhat)$ oscillation effect in \eqref{ToyModel3Result} require dedicated analyses to be performed by CMB experimental collaborations as the real on-the-sky variation of the polarization pattern on timescales short or comparable ($\sim$ a few hours to a few years) to the total observation times ($\sim$ months to years) demands consideration of non-public time-series data; such an analysis is beyond the intended scope of this work.

However, we can provide a rough, minimal estimate for the reach of such a search.
Most recently, Planck~\cite{Aghanim:2016fhp} has performed an analysis looking for an all-sky DC rotation of polarization that would give rise to a rotation of $E$ modes into $B$ modes, and have quoted null results (within uncertainties) for such searches that have statistical 68\% uncertainties at the $0.05^\circ$ level on the all-sky rotation angle~\cite{Aghanim:2016fhp}; see also~\citeR[s]{Hinshaw:2012aka,Kaufman:2013vbd} for older analyses with lower precision.
The systematic uncertainties on such searches are much larger, at the $0.3^\circ$ level~\cite{Aghanim:2016fhp}, but are dominated by uncertainty in the absolute calibration of the relative orientation of the bolometers~\cite{Aghanim:2016fhp} (but see, e.g.,~\citeR{Johnson:2015tga} for possible improvements in absolute calibration possible for ground-based detectors), which is irrelevant for an AC oscillation search so long as a fixed reference frame for the experiment can be maintained.

We estimate that a search in, e.g., the Planck time-series data for an oscillation signal should be able to resolve an oscillation with an amplitude roughly approaching the statistical power of the searches for these all-sky DC effects, although we have not considered in detail the impact of possible confounding systematic effects for such a search.
Based on this estimate, we assume a limit on the amplitude of the AC oscillation ($\gpgg\phi_0/2$) of $0.1^\circ$, at least, would be attainable at 95\% confidence in an analysis of currently existing time-series data.
This leads to
\begin{align}
\gpgg \phi_0 &\leq 3.5\times 10^{-3} \\
\Rightarrow \gpgg &\leq 1.6 \times 10^{-12}\,\text{GeV} ^{-1} \nl \times \lb( \frac{m_\phi}{10^{-21}\,\text{eV}}\rb) \times \lb(\kappa \times \frac{\rho_0}{ 0.3\, \text{GeV}/\text{cm}^3} \rb)^{-1/2},
\label{eq:ACest1}
\end{align}
which gives the diagonal dash-dotted (red) line in \figref{result}.
As an example of the possible reach of this technique, we consider the impact of improving the sensitivity to the AC oscillation amplitude to $0.01^\circ$ ($\gpgg\phi_0 \leq 3.5\times 10^{-4}$); since the amplitude of the rotation $\propto\gpgg $, this would correspond to the factor-of-10 improvement in the coupling reach compared to \eqref{ACest1} that is shown with the diagonal dotted (red) line in \figref{result}.
For comparison, the reach for the AC search exceeds the current (respectively, future) washout reach for a rotation amplitude of $0.059^\circ$ ($0.022^\circ$).

Note that with the minimal assumed sensitivity of the AC oscillation effect, both the washout and AC effects have similar reach, but since the AC oscillation effect is proportional to $\gpgg$, while the washout effect is proportional to $\sqrt{\gpgg}$ and is almost cosmic-variance limited already, the AC effect holds more promise for increased future reach.
In particular, we do not attempt to estimate the ultimate reach of this technique here, though it could possibly be significantly better than the reach shown in \figref{result}.
The minimal reach estimate we have used comes from what is already possible for measuring the static, absolute value of the polarization.
However, the effect we are looking for should be easier to measure since many of the limitations of the static search do not apply to a time-oscillating signal.  
Moreover, since the AC effect arises solely as a result of the local axion field value, the effect must be \emph{in phase} for photons arriving from any direction on the sky; this provides a highly non-trivial cross-check of any putative positive signal.

While we have based our reach estimates here on Planck results, searches for the AC effect are of course possible with time-series data from any of the existing or proposed ground-based, polarization-sensitive CMB experiments~\cite{Ahmed:2019zzz} (e.g., BICEP and the Keck Array~\cite{2014SPIE.9153E..1NA,2018SPIE10708E..2NK,Hui:2018cvg}, ACTpol~\cite{2016ApJS..227...21T}, SPTpol~\cite{McMahon_2009}, POLARBEAR-2~\cite{Inoue:2016jbg}, Simons Observatory~\cite{Ade:2018sbj}, etc.).
A dedicated experiment built to search for oscillating CMB polarization could also potentially have a significantly greater sensitivity to the AC effect than we plot in \figref{result}; we defer a detailed examination of this point to future work.

Aliasing and averaging effects will complicate setting experimental bounds for sufficiently short axion periods (high $m_\phi$) compared to, respectively, the interval between successive measurements of the same point on the sky, and the time for one observation during the overall integration time for any experiment.
Although aliasing effects render an experiment blind to certain specific axion masses (i.e., frequencies), they do not generally imply a loss of sensitivity at other masses; they are also mitigated against by the existence of multiple CMB polarization experiments which all have distinct survey strategies.
Moreover, with respect to the averaging effect, note that by `one observation' we do not mean the time required to integrate to get a decent statistical uncertainty on the polarization, but rather the much higher data acquisition rate of the experimental apparatus.
Given that the axion coherence time $\sim (m_\phi v^2)^{-1} \sim 10^6/m_\phi$ is much longer than the oscillation period $\sim 1/m_\phi$,%
\footnote{\label{ftnt:coherenceTime}%
        We exceed CAST bounds for masses below approximately $m_\phi \sim10^{-19}\,$eV, corresponding to frequencies $\sim 3\times 10^{-5}\,$Hz, or periods $\sim 10^{-3}$\,yr. 
        Thus, the coherence times for even the fastest-varying axions of interest to us are on the order of a thousand years.
    } %
it is possible through appropriate data analysis to extract an oscillatory AC signal even when the `single observation data points' are individually quite noisy.
However, as can be seen from \figref{result}, the relevant axion periods in the parameter range of interest to us are generally quite long, and certainly much longer than the data acquisition rates of modern CMB experiments; we have thus ignored complications from this effect in estimating the reach in our parameter range of interest.

\section{Discussion of previous work}
\label{sect:comparison}
In this section we distinguish the present paper from prior relevant work in the literature in some detail, supplementing the comments in this regard that we have already made in \sectref[s]{introduction} and \ref{sect:ExecutiveSummary}.

\citeR{Finelli:2008jv}%
\footnote{\label{ftnt:otherVersion}%
   	\citeR[s]{Galaverni:2009zz,Galaverni:2009gja} are related works by the same author(s), to which we omit further reference here for the sake of brevity.
    } %
derives the effects on the CMB of a time-varying axion field (see also~\citeR[s]{Liu:2006uh,Gubitosi:2009eu,Galaverni:2014gca}).
The authors of~\citeR{Finelli:2008jv} note and examine numerically that the usual static rotation approximation (i.e., the usual treatment of cosmic birefringence for cosmologically slowly varying fields) fails to accurately capture all relevant effects when the axion field oscillates rapidly.
In particular, the numerical comparison shown at Fig.~6 of~\citeR{Finelli:2008jv} between the full evolution and results based on the static rotation approximation does show an effect that appears to be the polarization washout effect we have examined in the present work.
However, while \citeR{Finelli:2008jv} remarks that this effect is clearly distinct from the results of a DC static birefringence analysis, that work does not present an analytical estimate for the magnitude of the effect, and no simple physical picture is developed to heuristically explain the origin of the washout.
Moreover, \citeR{Finelli:2008jv} does not utilize their numerical results including this effect to set limits on the axion--photon coupling parameter space. 
The present work discusses the washout effect in detail, and provides both a straightforward heuristic picture explaining its origin, and a simple analytical expression for its magnitude; we also use the effect to place new and powerful limits on the axion--photon coupling parameter space.

Moreover,~\citeR[s]{Finelli:2008jv,Galaverni:2009zz,Galaverni:2009gja} do mention that the linear polarization angle of the CMB would oscillate in time in an oscillating axion background [see, e.g., the discussion around Eq.~(90) of~\citeR{Finelli:2008jv}]; however, the results given in those works provide an incorrect estimate for the size of the effect, and they misidentify its origin.
In particular, those works give an estimate for the amplitude of this oscillation that is set by the axion field amplitude at decoupling;%
\footnote{\label{ftnt:localAxionField}%
   	Moreover, although they drop the local axion field amplitude as small in making this estimate, the results at the first line of Eq.~(90) of~\citeR{Finelli:2008jv} indicate that the authors of that work did not account for the the local overdensity of the axion dark matter inside the Milky Way in estimating the axion field amplitude at late times.
          } %
we have shown that the early-universe axion field oscillations average out to zero, instead giving rise to the polarization washout effect.
We have also shown that, as a result, it is the local, late-time axion field that controls the amplitude of the AC effect.
This distinction is qualitative as well as quantitative: the understanding developed in the present work that the AC oscillation arises from the local, late-time value of the axion field implies that the oscillation signal is in phase in different detectors looking at different directions on the sky, which is to our knowledge a novel observation.

Separately, a number of recent works (e.g.,~\citeR{Sigl:2018fba} in the CMB context) have treated the net rotation angle accumulated by a photon traversing $N$ different coherently oscillating axion DM patches by summing incoherently over the mean-square rotation angles from each patch, obtaining a `random-walk' $\sqrt{N}$ enhancement.
As we have shown, such a treatment is erroneous, as the net rotation is simply proportional to the difference of axion field values at photon emission and photon absorption, independent of the intervening behavior of the axion field~\cite{Harari:1992ea} (see also \appref{WKB} for a discussion of the assumptions underlying this result).%
\footnote{\label{ftnt:note_added}%
		 See also the `note added' in~\citeR{Ivanov:2018byi}. 
		 Similar erroneous treatments also appeared in the \texttt{arXiv v1} preprints of~\citeR[s]{Liu:2019brz,Caputo:2019tms,Fujita:2018zaj}, but were corrected in revised versions (\texttt{arXiv v2}, and/or published) either as the present work was being finalized \cite{Liu:2019brz,Caputo:2019tms}, or after the \texttt{arXiv v1} preprint of the present work appeared \cite{Fujita:2018zaj}.
	} %

One other related previous work of which we are aware,~\citeR{Liu:2016dcg}, looks for static anisotropic birefringence induced by axion dark matter at the time of last scattering; however, as we have discussed, axion dark matter does not actually exhibit an unsuppressed signal of this type, as the net rotation angle induced at the last scattering surface averages to approximately zero point-by-point on the sky. 

Our bounds compare favorably with some recent analyses of polarization rotation induced by axion dark matter in astrophysical contexts: over the relevant range of masses, we exceed by a factor of a few the limits set in~\citeR{Ivanov:2018byi} from considering active galactic nuclei (AGNs) as the polarized source;%
\footnote{\label{ftnt:AGN_DM}%
		 Although we note that results of~\citeR{Ivanov:2018byi} require knowledge of the dark-matter density near the center of elliptical galaxies hosting such AGNs, which is naturally subject to some considerable uncertainty.
	} %
our results also exceed the recently revised (see footnote \ref{ftnt:note_added}) limits and projections made in~\citeR{Fujita:2018zaj}, using a protoplanetary disk as the polarized source; and our limits are comparable to the recently revised (see footnote \ref{ftnt:note_added}) projections made in~\citeR{Liu:2019brz}, and limits set in~\citeR{Caputo:2019tms},%
\footnote{\label{ftnt:coherence_length}%
		 The pulsar considered in~\citeR{Caputo:2019tms}, J0437-4715, is only $D\sim 156$\,pc distant from Earth~\cite{Deller:2008jx}, which is within an axion coherence length $\lambda_{\text{coh.}}\sim 2\pi / (m_a v_0) \sim 156\,\text{pc} \times (3.5\times 10^{-22}\,\text{eV}/m_a) \times (220\,\text{km}\,\text{s}^{-1}/v_0)$ for some part of the axion mass range of interest, and close enough that the axion field amplitude (i.e., DM density) at the pulsar would be similar to that at Earth.
		 But then the difference in the axion field between emission at $x'=(t-D,D\nhat)$ and absorption at $x=(t,\bm{0})$ is $\Delta \phi \approx \phi_0 \lb[ \cos\lb(m_a t\rb) - \cos\lb(m_a (t-D) + \delta \rb)\rb]$, where $\delta \sim \mathcal{O}(\pi D/\lambda_{\text{coh.}})$.
    	 We would thus expect to see some loss of sensitivity in the results of~\citeR{Caputo:2019tms} at certain specific axion masses [$(m_a D-\delta)\! \mod 2\pi \approx 0$] when $m_a \ll 3.5\times 10^{-22}\,$eV.
	} %
using pulsars as the linearly polarized source.

In summary, our work is to our knowledge the first to identify the AC oscillation effect in the CMB context as arising from the local axion field,%
\footnote{\label{ftnt:otherworks}%
		In addition to~\citeR[s]{Finelli:2008jv,Galaverni:2009zz,Galaverni:2009gja}, see also~\citeR[s]{Fujita:2018zaj,Ivanov:2018byi,Liu:2019brz,Caputo:2019tms} for recent analyses in astrophysical contexts employing the idea of an oscillatory effect.
	} %
and is thus the first to provide both a correct estimate for the size of the effect, and to point out that the oscillations would be in phase in different CMB detectors observing photons arriving from any direction on the sky, which allows for non-trivial cross-checks of any putative positive signal.
Our work is also to our knowledge the first to give a simple derivation and physical explanation of the polarization washout effect, and use it to bound the axion--photon coupling.

Moreover, we urge caution in utilizing previous CMB-based birefringence bounds on the axion dark-matter parameter space that have not considered the effects of the rapid time variation of the field at the decoupling epoch.

\section{Conclusion}
\label{sect:Conclusion}

In this work, we proposed a new technique to search for axion dark matter in the lowest allowed mass range.
In particular, this appears to be one of the most sensitive ways to directly detect (theoretically well-motivated) axion-type fuzzy dark matter~\cite{Hui:2016ltb}.
We found that axion dark matter has two novel effects on the polarization of the CMB.
First, a uniform reduction in polarization power from the standard $\Lambda$CDM expectation for all $l \gtrsim 20$.
This arises from the axion field oscillating many times during the CMB decoupling epoch, resulting in a washout of imprinted polarization as compared to the net polarization expected to be imprinted by the local CMB temperature quadrupole polarization source at last scattering.
Second, a real on-the-sky AC oscillation of the CMB polarization at the period of the axion field, which is amenable to experimental detection, but which requires detailed and dedicated time-series analyses by CMB experimental collaborations.

The washout effect is quadratic in a small number, the axion field times the axion--photon coupling ($\sim \gpgg^2 \phi^2$), but arises from early times when the axion field is larger.
The AC effect is linear in the axion field times the coupling ($\sim \gpgg \phi$), but arises from late times when the axion field is smaller.
Given the current and projected sensitivities as shown in our main results in \figref{result}, these two effects are seen to be of roughly comparable reach in constraining axion parameter space in the fuzzy-dark-matter region $m_\phi \sim 10^{-19}$--$10^{-22}$\,eV.
Indeed, we have used the washout effect to set a bound on axion dark matter using current Planck results that is several orders of magnitude beyond previous limits.
This bound is already close to the ultimate cosmic-variance-limited reach for this effect; see \figref{result}.

Beyond setting limits, both effects have discovery potential.  No CMB physics can mimic either effect, and the AC effect especially appears distinct from any cosmological background.
Of course, as in any precision experiment, care must be taken to eliminate other backgrounds, but there are strong checks on a positive signal.
For example, all CMB detectors must see the same effect; in particular, the signal for the AC effect must be \emph{in phase} in the different detectors, independent of their location on or near Earth, and independent of the location on the sky they observe.
This would be dramatic confirmation of a detection.
For the future, the AC effect, though requiring a dedicated analysis, has the greater potential reach since it is linear in the coupling (and not limited by cosmic variance).
We have drawn two representative curves in \figref{result} for the AC analysis, but we suspect that the ultimate reach of this method could even be beyond either of these.
Thus, undertaking these AC analyses is important as this is ultimately the better way to search for axion dark matter in the lowest mass ranges.

In general, there are multiple ways to directly search for axion dark matter in the lighter part of its mass range, both with terrestrial experiments (e.g.,~\citeR[s]{Abel:2017rtm, Graham:2017ivz, Terrano:2019clh, Wu:2019exd, Garcon:2019inh, Alonso:2018dxy, Safronova:2017xyt}) and astrophysical observations (e.g.,~\citeR[s]{Ivanov:2018byi,Liu:2019brz,Caputo:2019tms}).
CMB polarization appears to be one of the most promising approaches for the very lightest end of the axion mass range, and offers the exciting possibility of directly detecting fuzzy dark matter.

\acknowledgments
We thank Zeeshan Ahmed, Simone Ferraro, Kent Irwin, Uro\u s Seljak, and Leonardo Senatore for useful conversations. 
This work was supported by the Heising-Simons Foundation Grants No.~2015-037, No.~2015-038, and No.~2018-0765, and DOE HEP QuantISED Award No.~100495.
M.A.F.~and P.W.G.~are further supported by DOE Grant No.~DE-SC0012012, NSF Grant No.~PHY-1720397, and the Gordon and Betty Moore Foundation Grant No.~GBMF7946. 
S.R.~was supported in part by the NSF under Grants No.~PHY-1638509 and No.~PHY-1507160, and the Simons Foundation Award No.~378243.

\appendix
\section{Derivation of the polarization rotation effect}
\label{app:WKB}
\newcommand{\deta}{\partial_\eta}
\newcommand{\dz}{\partial_z}

The classical field equation for the axion field arising from \eqref{GRaction}, assuming that $V(\phi) \equiv \frac{1}{2} m_\phi^2 \phi^2$, is
\begin{align}
\Box \phi + m_\phi^2 \phi &= - \frac{\gpgg}{4} F_{\mu\nu}\widetilde{F}^{\mu\nu}.
\label{eq:ALPEOM}
\end{align}
Throughout, we ignore the backreaction term on the RHS; we justify this quantitatively at the end of this Appendix.
The classical field equations for the photon are~\cite{Wilczek:1987mv}
\begin{align}
\nabla_\mu F^{\mu\nu} &= J^\nu - \gpgg (\nabla_\mu \phi) \widetilde{F}^{\mu\nu}.
\label{eq:FLRW_F_eom}
\end{align}
In Lorenz gauge (generalized to curved spacetime), $g^{\alpha\beta}\nabla_\alpha A_\beta = 0$, the field equations for the potentials in a source-free region ($J^\nu = 0$) become%
\footnote{\label{ftnt:abuseofnotation}%
		 We acknowledge an abuse of notation on the RHS of \eqref{AEoM}: $\gpgg$ is the axion--photon coupling; $g_{\nu\alpha}$ is the metric.
	} %
\begin{align}
\Box A_\nu + {R^{\mu}}_\nu A_\mu &= - g_{\nu \alpha}  (\gpgg \partial_{\mu}\phi ) \epsilon^{\mu\alpha\lambda\rho} \partial_\lambda A_\rho,
\label{eq:AEoM}
\end{align}
where we have used that $\epsilon^{\mu\alpha\lambda\rho} \nabla_\lambda A_\rho = \epsilon^{\mu\alpha\lambda\rho} \partial_\lambda A_\rho$ owing to the symmetries of the Levi-Civita tensor and Christoffel symbols.

We specialize to a homogeneous, isotropic FLRW universe with scale factor $a$, and work in the conformal--comoving co-ordinate system $(\eta, \bm{x})$ such that the line element is $ds^2 = [a(\eta)]^2 \lb( d\eta^2 - d\bm{x}^2 \rb)$; $\eta$ is conformal time.
As a specific case, we will consider $\phi = \phi(\eta,z)$, and seek solutions to the photon equations of motion which take the form $A_\mu = A_\mu(\eta,z)$ (i.e., solutions in which the variation in the fields is negligible in the direction transverse to some selected direction); see also \citeR{Harari:1992ea} for a slightly more general discussion given in terms of the $\bm{E}$ and $\bm{B}$ fields.
Under these assumptions, a transverse solution consistent with the $\nu=0,3$ photon equations of motion and the Lorenz gauge condition is $A_0=A_3=0$, while the $\nu = 1,2$ equations of motion can be decoupled by defining the definite-helicity transverse field variables as at \eqref{AsigmaDefn}, in terms of which we have
\begin{align}
\partial_\eta^2 A_\sigma - \partial_z^2 A_\sigma = i\sigma \gpgg \lb[ (\partial_z \phi) (\partial_\eta A_\sigma ) - (\partial_\eta \phi)(\partial_z A_\sigma) \rb].
\label{eq:eomFLRW}
\end{align}

We follow a WKB-like perturbative solution approach.
Consider the following solution ansatz for \eqref{eomFLRW}:
\begin{align}
A_\sigma(\eta,z) &= F_\sigma(\eta,z) \nl \times \exp\Big[\! -i \omega (\eta-\eta') + ik(z-z') \nl\qquad\qquad + iG_\sigma(\eta,z;\eta',z') \Big],
\label{eq:WKBA}
\end{align}
where $F_\sigma$ and $G_\sigma$ are real functions. 
Substituting into \eqref{eomFLRW}, we have
\begin{widetext}
\begin{align}
&\lb[ \deta^2 \ln F_\sigma + \lb(\deta \ln F_\sigma \rb)^2 - \omega^2 - (\deta G_\sigma)^2 + 2\omega \deta G_\sigma \rb] + i \lb[  - 2 ( \omega - \deta G_\sigma ) \deta \ln F_\sigma +\deta^2 G_\sigma \rb] \nl
+\lb[ -\dz^2 \ln F_\sigma - \lb(\dz \ln F_\sigma \rb)^2 + k^2 + (\dz G_\sigma)^2 + 2k\dz G_\sigma \rb] + i \lb[  - 2 ( k+ \dz G_\sigma) \dz \ln F_\sigma - \dz^2 G_\sigma \rb]  \nl
=  \sigma \gpgg (\dz \phi)\lb[ i \deta \ln F_\sigma + \omega - \deta G_\sigma \rb]  + \sigma \gpgg (\deta \phi)\lb[ -  i \dz \ln F_\sigma + k + \dz G_\sigma \rb]  \label{eq:WKBEoM1}
\end{align}
Equating the real and imaginary parts, we have, using that $\partial_x^2 \ln A + ( \partial_x \ln A)^2 = \partial_x^2A / A$ for $x\in\{\eta,z\}$,
\begin{align}
& k^2 - \omega^2 + 2\omega \deta G_\sigma + 2k\dz G_\sigma + \frac{ \deta^2 F_\sigma}{F_\sigma} - \frac{ \dz^2 F_\sigma }{ F_\sigma } + (\dz G_\sigma)^2  - (\deta G_\sigma)^2  \nl
=  \sigma \gpgg (\dz \phi)\lb[ \omega - \deta G_\sigma \rb]  +  \sigma \gpgg(\deta \phi)\lb[  k + \dz G_\sigma \rb]  \label{eq:WKBeqnfull1} \\[5ex]
& - 2 ( \omega - \deta G_\sigma) \deta \ln F_\sigma +\deta^2 G_\sigma  - 2 ( k+ \dz G_\sigma) \dz \ln F_\sigma - \dz^2 G_\sigma  \nl
=  \sigma \gpgg (\dz \phi)\lb[  \deta \ln F_\sigma  \rb]  +  \sigma \gpgg (\deta \phi)\lb[ -  \dz \ln F_\sigma  \rb]  \label{eq:WKBeqnfull2}.
\end{align}
\end{widetext}

In order to make progress, we introduce a formal small parameter $\epsilon \ll 1$ (we quantify this below) by assuming that $\phi$, $F_\sigma$, and $G_\sigma$ are functions of `slow' time (space) variables $\tau$ ($\xi$) which are considered to vary slowly on the scale of $1/\omega$ ($1/k$) (the reader will recognize this as an alternative technique for the derivation of the WKB approximation): 
\begin{align}
\tau &\equiv \epsilon \, \omega \eta \Rightarrow \deta = \epsilon\, \omega \partial_\tau \\
\xi &\equiv \epsilon \, k z \Rightarrow \partial_z = \epsilon\, k \partial_\xi.
\label{eq:slowVariables}
\end{align}
Substituting into \eqref[s]{WKBeqnfull1} and (\ref{eq:WKBeqnfull2}), we obtain
\begin{widetext}
\begin{align}
&n_0^2 - 1 + 2 \epsilon  \partial_\tau G_\sigma + 2\epsilon n_0^2 \partial_\xi  G_\sigma + \epsilon^2 \frac{  \partial_\tau^2 F_\sigma}{F_\sigma} - \epsilon^2 n_0^2 \frac{ \partial_\xi^2 F_\sigma }{ F_\sigma } + \epsilon^2 n_0^2  (\partial_\xi G_\sigma)^2  - \epsilon^2 (\partial_\tau G_\sigma)^2  \nl
= \epsilon n_0 \sigma \gpgg ( \partial_\xi \phi)\lb[ 1 -  \epsilon \partial_\tau G_\sigma \rb]  + \epsilon n_0 \sigma \gpgg ( \partial_\tau \phi)\lb[  1 + \epsilon \partial_\xi G_\sigma \rb]  \label{eq:WKBeqnfull1b} \\[5ex]
& - 2 \epsilon  ( 1 -  \epsilon \partial_\tau G_\sigma)   \partial_\tau \ln F_\sigma + \epsilon^2 \partial_\tau^2 G_\sigma  - 2 n_0^2 \epsilon ( 1+ \epsilon \partial_\xi G_\sigma) \partial_\xi \ln F_\sigma -  \epsilon^2 n_0^2 \partial_\xi^2 G_\sigma  \nl
= \epsilon^2  n_0 \sigma \gpgg ( \partial_\xi \phi)\lb[   \partial_\tau \ln F_\sigma  \rb]  - \epsilon^2 n_0 \sigma \gpgg ( \partial_\tau \phi)\lb[ \partial_\xi \ln F_\sigma  \rb]  \label{eq:WKBeqnfull2b}.
\end{align}
\end{widetext}
where we have defined $n_0 \equiv k/\omega> 0$.
Equating equal powers of $\epsilon$ in these equations independently, it follows that, up to $\mathcal{O}(\epsilon)$, we have 
\begin{align}
n_0^2 &= 1, \label{eq:10}\\
\partial_\tau G_\sigma + n_0^2 \partial_\xi G_\sigma &= n_0 (\sigma \gpgg/2) ( \partial_\xi \phi +  \partial_\tau \phi )  \label{eq:11}\\
\partial_\tau \ln F_\sigma + n_0^2 \partial_\xi \ln F_\sigma &= 0 \label{eq:21},
\end{align}
where \eqref{10} and \eqref{11} arise respectively from the $\epsilon^0$ and $\epsilon^1$ terms in \eqref{WKBeqnfull1b}, and \eqref{21} arises from the $\epsilon^1$ terms in \eqref{WKBeqnfull2b}.
We may ignore the equations at second and higher order, as $\epsilon \ll 1$ renders these corrections irrelevant (see below for a numerical estimate of $\epsilon$).
We thus have $n_0 = 1$, which implies that
\begin{align}
( \partial_\tau + \partial_\xi ) \lb( G_\sigma -  \sigma \frac{ \gpgg}{2} \phi \rb) &= 0   \\
(\partial_\tau + \partial_\xi ) \ln F_\sigma  &= 0.
\label{eq:FGeom}
\end{align}
The solutions to $(\partial_\tau + \partial_\xi ) h(\tau,\xi) = 0$ are any functions $h(\tau,\xi) = h(\tau-\xi)$, so we may write without loss of generality
\begin{align}
G_\sigma(\tau,\xi) &= \sigma \frac{ \gpgg}{2}\phi(\tau,\xi)  + h_1(\tau-\xi)   \\
F_\sigma(\tau,\xi)  &= h_2(\tau-\xi),
\label{eq:FGsoln1}
\end{align}
where $h_{1,2}$ are general real functions to be determined.
At this point, having derived the approximate solution, there is no purpose served by continuing to distinguish $h_{1,2}$ as being functions of $(\tau,\xi)$; we thus revert to using $(\eta,z)$ as the arguments of these functions.

If we demand that in the limit $\gpgg=0$ we have the standard plane-wave solution
\begin{align}
A_\sigma(\eta,z) = \bar{A}_\sigma e^{-i\omega(\eta-\eta')+ik(z-z')},
\label{eq:Asoln1}
\end{align}
with $\bar{A}_\sigma = A_\sigma(\eta',z')$ a constant, this can clearly be achieved by setting
\begin{align}
h_2(\eta-z) &= |\bar{A}_\sigma| = \text{const.} &&(\gpgg=0),\\
h_1(\eta-z) &= \arg\lb[ \bar{A}_\sigma \rb] = \text{const.} && (\gpgg=0),
\label{eq:h1h2soln1}
\end{align}
which uniquely defines $h_{1,2}$ as constant functions.
If we require that for $\gpgg\neq0$ we have a plane-wave solution with the same normalization and phase at the source as for the $\gpgg = 0$ case,
\begin{align}
A_\sigma(\eta',z') = \bar{A}_\sigma,
\label{eq:AIC}
\end{align}
this can be achieved by the choice 
\begin{align}
h_2(\eta-z) &= |\bar{A}_\sigma| = \text{const.},\\
h_1(\eta-z) &= \arg\lb[ \bar{A}_\sigma \rb] - \sigma (\gpgg /2) \phi(\eta',z') = \text{const.};
\label{eq:h1h2soln2}
\end{align}
\eqref{PhaseShift} follows immediately. 
Note that we have \emph{not} assumed that $\gpgg\phi$ is small in the above derivation, only that $\phi$ varies slowly on the time or length scales of the photon field (see also \citeR{Harari:1992ea} for a similar observation in the context of an alternative derivation).

In our work, the numerical estimate for the small parameter $\epsilon$ is given approximately by $\epsilon \sim m_\phi / \omega$ [as the axion field in our work plays the role of non-relativistic dark matter, its temporal gradients $|\dot{\phi}|\sim m|\phi|$ are $\gtrsim \mathcal{O}(10^3)$ larger than the spatial gradients, $|\nabla \phi| \sim mv|\phi|$, so the ratio of temporal evolution timescales is the relevant small parameter].
Since we are concerned with the CMB, which has a peak frequency around $f \sim 160\,\text{GHz}$, we have $\omega \sim  0.7 \,\text{meV}$, and so $\epsilon \sim 10^{-19} ( m_\phi / 10^{-22}\ \text{eV})$, which is vanishingly small throughout our region of interest; we are thus justified in keeping only correction terms arising from the equations which are first order in $\epsilon$, which are precisely the terms shown at \eqref{PhaseShift}.

Finally, we return to the justification for dropping the backreaction term on the RHS of \eqref{ALPEOM}. 
First, note that $-(\gpgg/4)F\tilde{F} = \gpgg \bm{E}\cdot\bm{B}$. 
Our conclusion that the polarization simply rotates at leading order immediately implies that, at leading order, we would still have $\bm{E} \cdot \bm{B} = 0$ for any transverse wave propagating in the axion background (absent any electromagnetic background); this alone would justify the neglect of the backreaction term arising from the passage of CMB photons through the axion background at leading order.

However, suppose that there is a sub-leading term whose impact we have ignored in the main analysis, but which might cause $\bm{E} \cdot \bm{B}=0$ to fail at $\mathcal{O}(\gpgg)$.
Since any such effect must vanish in the limit of constant $\phi$, a very conservative estimate for the maximum magnitude of the backreaction arising from a transverse light wave of frequency $\omega$ can be given as $|\gpgg \bm{E}\cdot\bm{B}| \lesssim \gpgg \bm{E}^2 \times \gpgg |\dot{\phi}|/\omega \sim \rho_R \gpgg^2 |\dot{\phi}| / \omega \sim \rho_R \gpgg^2 m_\phi |\phi| / \omega$, where in the last step we approximated $|\dot{\phi}| \sim m_\phi |\phi|$, in accordance with the assumption of vanishing backreaction at leading order; we also assumed that the axion field has larger temporal than spatial gradients (i.e., is non-relativistic).
Furthermore, in the first step, we took the conservative estimate $\bm{E}\cdot\bm{B} \sim \bm{E}^2 \sim \rho_R$, the full radiation energy density.

In order to neglect the backreaction, the size of the term $|\gpgg \bm{E}\cdot\bm{B}|$ on the RHS of \eqref{ALPEOM} is required to be much less than the individual terms on the LHS of \eqref{ALPEOM}; under the assumptions just discussed, these terms are of size $3H \dot{\phi}$, and $m_\phi^2 \phi$.
The most restrictive condition which must be imposed to neglect the backreaction term in \eqref{ALPEOM} is $|\gpgg \bm{E}\cdot\bm{B}| \ll 3H |\dot{\phi}|$.
The less restrictive condition, $|\gpgg \bm{E}\cdot\bm{B}| \ll m_\phi^2 |\phi|$, is then necessarily satisfied because the axion field is the cold dark matter, and must thus necessarily be oscillating, which imposes the condition $3H < m_\phi$, or $|3H\dot{\phi}| < |m_\phi^2\phi|$, assuming consistently that $|\dot\phi|\sim m_\phi|\phi|$.
More explicitly, the largest $H$ which is strictly relevant for the axion dark matter is the value attained around matter--radiation equality, when $H \sim T^2_{\text{eq.}}/M_{\text{Pl.}} \sim 10^{-28}\,$eV (taking $T_{\text{eq.}} \sim \text{eV}$); throughout the entire range of masses $m_\phi \gtrsim 10^{-22}\,$eV in which the axion field can be all the dark matter consistent with small-scale structure limits, we thus have $3H \ll m_\phi$ at all times from matter--radiation equality to the present day.
In particular, the condition $3H < m_\phi$ is easily satisfied throughout the mass range where our limits are competitive with or exceed those of CAST: i.e., $m_\phi \sim 10^{-22}\,\text{eV}$ to $10^{-18}\,$eV; see \figref{result}. 

We thus require $|\gpgg \bm{E}\cdot\bm{B}| \ll 3H |\dot{\phi}|$ to ignore the backreaction in \eqref{ALPEOM}.
Using the estimate developed above that $|\gpgg \bm{E}\cdot\bm{B}| \lesssim  \rho_R \gpgg^2 |\dot{\phi}| / \omega$, this is achieved whenever $|3H \dot{\phi}| \gg |\rho_R \gpgg^2 \dot{\phi} / \omega|$, or $\rho_R \gpgg^2 /( \omega H) \ll 1$.
At the order of magnitude level, it is easily seen that this is satisfied by a large margin. 
After matter--radiation equality, we estimate $H^2 \sim \rho_{\text{tot.}}/M_{\text{Pl.}}^2 \sim T^3 T_{\text{eq.}} M_{\text{Pl.}}^{-2}$, while $\rho_R \sim T^4$, and $\omega \sim T$ at the CMB spectral peak.
Therefore, we require that $\gpgg^2  M_{\text{Pl.}} T(T/T_{\text{eq.}})^{1/2} \ll 1$.
Conservatively taking $\gpgg \sim 10^{-10}\,\text{GeV}^{-1}$ to be slightly larger than the largest value allowed by CAST limits \cite{Anastassopoulos:2017ftl} (see \figref{result}), and conservatively taking the largest relevant temperature, $T\sim T_{\text{eq.}} \sim \text{eV}$, we find that $ \gpgg^2  M_{\text{Pl.}} T (T/T_{\text{eq.}})^{1/2} \sim 10^{-10} \ll 1$.

We thus conclude that, subject to the assumptions detailed in this Appendix, our results in principle safely hold throughout the mass range $10^{-22}\,\text{eV}\lesssim m_\phi \ll \omega_{\textsc{cmb}} \sim 10^{-3}\,\text{eV}$, for all couplings $\gpgg$ around or below current CAST bounds.

\section{Stokes parameters}
\label{app:stokes}

\subsection{Classical definitions}
\label{app:stokesClassical}
We define the Stokes parameters following the conventions of~\citeR{Jackson}.
To translate these results to the axis conventions of \figref{axes}, replace $\bm{\hat{e}_1}\rightarrow \etheta$ and $\bm{\hat{e}_2}\rightarrow \ephi$.
Suppose that the electric field can be written as (the real part of)
\begin{align}
\bm{E} &\equiv e^{-ikx} \lb[ E^1 \bm{\hat{e}_1} + E^2 \bm{\hat{e}_2} \rb] \\
&\equiv e^{-ikx} \lb[  E^+ \bm{\hat{e}_+} + E^- \bm{\hat{e}_-} \rb], 
\label{eq:Field1}
\end{align}
where
\begin{align}
\bm{\hat{e}_\pm} &\equiv \frac{1}{\sqrt{2}}\lb( \bm{\hat{e}_1} \pm i \bm{\hat{e}_2} \rb),\\
E^\pm &\equiv \frac{1}{\sqrt{2}} \lb( E^1 \mp i E^2 \rb).
\label{eq:Field2}
\end{align}
The Stokes parameters $(I,Q,U,V)$ defined with respect to these frames are
\begin{align}
I &\equiv | \bm{\hat{e}_1} \cdot \bm{E} |^2 + | \bm{\hat{e}_2} \cdot \bm{E} |^2 \\ &= |E^1|^2 + |E^2|^2 \\
	&\equiv | \bm{\hat{e}_+}^* \cdot \bm{E} |^2 +| \bm{\hat{e}_-}^* \cdot \bm{E} |^2 \\&= |E^+|^2 + |E^-|^2
\end{align}
\begin{align}
Q &\equiv  | \bm{\hat{e}_1} \cdot \bm{E} |^2 - | \bm{\hat{e}_2} \cdot \bm{E} |^2 \\&= |E^1|^2 - |E^2|^2 \\
	&\equiv 2\, \text{Re}\,\lb[ \lb( \bm{\hat{e}_+}^* \cdot \bm{E} \rb)^* \lb( \bm{\hat{e}_-}^* \cdot \bm{E} \rb) \rb] \\&= 2 |E^+||E^-| \cos\lb( \alpha_- - \alpha_+ \rb),\\[2ex]
U &\equiv 2\, \text{Re}\,\lb[ \lb( \bm{\hat{e}_1}^* \cdot \bm{E} \rb)^* \lb( \bm{\hat{e}_2}^* \cdot \bm{E} \rb) \rb] \\&= 2 |E^1||E^2| \cos\lb( \alpha_2 - \alpha_1 \rb) \\
	&\equiv 2\, \text{Im}\,\lb[ \lb( \bm{\hat{e}_+}^* \cdot \bm{E} \rb)^* \lb( \bm{\hat{e}_-}^* \cdot \bm{E} \rb) \rb] \\&= 2 |E^+||E^-| \sin\lb( \alpha_- - \alpha_+ \rb) \\[2ex]
V &\equiv 2\, \text{Im}\,\lb[ \lb( \bm{\hat{e}_1}^* \cdot \bm{E} \rb)^* \lb( \bm{\hat{e}_2}^* \cdot \bm{E} \rb) \rb] \\&= 2 |E^1||E^2| \sin\lb( \alpha_2 - \alpha_1 \rb) \\
	&\equiv  | \bm{\hat{e}_+}^* \cdot \bm{E} |^2 - | \bm{\hat{e}_-}^* \cdot \bm{E} |^2 \\&= |E^+|^2 - |E^-|^2 
\label{eq:Stokes}
\end{align}
where $\varphi_X$ is defined by $E^X \equiv |E^X| e^{i\alpha_X}$ for any $X \in \{ 1,2,+,-\}$.

A field which is linearly polarized has $V=0$; $V\neq0$ implies elliptical polarization, with $V = \pm I$ implying circular polarization of helicity $\pm 1$. 
A linearly polarized field with $Q>0, U=0$ is polarized along the $\bm{\hat{e}_1}$-axis ($E^1\neq 0$ and $E^2 = 0$); whereas a linearly polarized field with $Q<0, U=0$ is polarized along the $\bm{\hat{e}_2}$-axis ($E^2\neq 0$ and $E^1 = 0$).
On the other hand, a linearly polarized field with $U>0, Q=0$ is polarized along the $(\bm{\hat{e}_1}+\bm{\hat{e}_2})$-axis ($E^1=E^2$); whereas a linearly polarized field with $U<0, Q=0$ is polarized along the $(\bm{\hat{e}_1}-\bm{\hat{e}_2})$-axis ($E^1=-E^2$).

\subsection{Quantum mechanical definitions}
\label{app:stokesQuantum}
As explained in~\citeR{Kosowsky:1994cy}, we may equivalently define the Stokes parameters in terms of quantum mechanical operators acting on photon polarization states.
Using the ($\etheta$, $\ephi$) basis, we define the operators
\newcommand{\ethetabra}[1][]{\langle\bm{\hat{e}_{\vartheta}}(\bm{\hat{n}^{#1}})|}
\newcommand{\ephibra}[1][]{\langle\bm{\hat{e}_{\varphi}}(\bm{\hat{n}^{#1}})|}
\newcommand{\eonebra}[1][]{\langle\bm{\hat{e}_{1}}(\bm{\hat{n}^{#1}})|}
\newcommand{\etwobra}[1][]{\langle\bm{\hat{e}_{2}}(\bm{\hat{n}^{#1}})|}
\begin{align}
\hat{I} &\equiv \ethetaket\ethetabra +  \ephiket\ephibra \\
\hat{Q} &\equiv \ethetaket\ethetabra -  \ephiket\ephibra \\
\hat{U} &\equiv \ethetaket\ephibra +  \ethetaket\ephibra \\
\hat{V} &\equiv i \ephiket\ethetabra - i \ethetaket\ephibra.
\label{eq:StokesQM}
\end{align}
A single-photon state linearly polarized along the direction $\bm{\psi} \equiv \cos\psi \etheta + \sin\phi \ephi$ has the polarization state $|\bm{\psi}\rangle = \cos\psi \ethetaket + \sin\psi\ephiket$, and thus has
\begin{align}
I_\psi &\equiv \langle\bm{\psi}| \hat{I} | \bm{\psi} \rangle = \cos^2\psi + \sin^2\psi = 1\\
Q_\psi &\equiv \langle\bm{\psi}| \hat{Q} | \bm{\psi} \rangle = \cos^2\psi - \sin^2\psi = \cos(2\psi)\\
U_\psi &\equiv \langle\bm{\psi}| \hat{U} | \bm{\psi} \rangle = 2\cos\psi \sin\psi = \sin(2\psi)\\
V_\psi &\equiv \langle\bm{\psi}| \hat{V} | \bm{\psi} \rangle = 0,
\label{eq:StokesQMexample}
\end{align}
which indicates that a single linearly polarized photon is present, and that $Q_\psi \pm i U_\psi = e^{\pm 2i\psi}$.
With appropriate scaling factors, it is clear that these definitions agree with the definitions for the Stokes parameters in terms of the classical electric field of an electromagnetic wave given in \appref{stokesClassical}.

\bibliographystyle{apsrev4-2}
\bibliography{main.bib}

\end{document}